\documentclass[aps,prx,floatfix,twocolumn,longbibliography,nofootinbib]{revtex4-2}

\usepackage[utf8]{inputenc}
\usepackage[T1]{fontenc}
\usepackage{physics}
\usepackage{graphicx}
\usepackage{adjustbox}
\usepackage{amsmath, amsfonts, amssymb}
\usepackage[breaklinks=true, colorlinks]{hyperref}
\hypersetup{linkcolor=red, citecolor=blue}

\usepackage{bbm, bm}
\usepackage[dvipsnames]{xcolor}
\usepackage{makecell}
\usepackage{upgreek}
\usepackage{nicefrac}

\usepackage{tikz}
\usepackage{float}
\usepackage{tikz-cd}
\usepackage{quantikz}

\makeatletter
\newcommand*{\encircled}[1]{\relax\ifmmode\mathpalette\@encircled@math{#1}\else\@encircled{#1}\fi}
\newcommand*{\@encircled@math}[2]{\@encircled{$\m@th#1#2$}}
\newcommand*{\@encircled}[1]{%
  \tikz[baseline,anchor=base]{\node[draw,circle,outer sep=0pt,inner sep=.2ex] {#1};}}
\makeatother

\newcommand{\OO}{\mathcal{O}}
\newcommand{\Var}[0]{{ \text{Var}  }}
\newcommand{\diag}[0]{{ \text{diag} }}
\newcommand{\xx}[0]{{ \vb{x} }}

\DeclareMathOperator*{\argmin}{argmin}

\begin{document}

\title{Neural-network quantum states for many-body physics}

\author{Matija Medvidović}
\email{matija.medvidovic@columbia.edu}
\affiliation{Center for Computational Quantum Physics, Flatiron Institute, 162 5th Avenue, New York, NY 10010, USA}
\affiliation{Department of Physics, Columbia University, NY 10027, USA}

\author{Javier Robledo Moreno}
\email{j.robledomoreno@ibm.com}
\affiliation{Center for Computational Quantum Physics, Flatiron Institute, 162 5th Avenue, New York, NY 10010, USA}
\affiliation{Center for Quantum Phenomena, Department of Physics, New York University, 726 Broadway, New York, NY 10003, USA}
\affiliation{IBM Quantum, IBM Research - 1101 Kitchawan Rd, Yorktown Heights, NY 10598, USA}

\date{\today}

\begin{abstract}
    Variational quantum calculations have borrowed many tools and algorithms from the machine learning community in the recent years. Leveraging great expressive power and efficient gradient-based optimization, researchers have shown that trial states inspired by deep learning problems can accurately model many-body correlated phenomena in spin, fermionic and qubit systems. In this review, we derive the central equations of different flavors variational Monte Carlo (VMC) approaches, including ground state search, time evolution and overlap optimization, and discuss data-driven tasks like quantum state tomography. An emphasis is put on the geometry of the variational manifold as well as bottlenecks in practical implementations. An overview of recent results of first-principles ground-state and real-time calculations is provided.
\end{abstract}

\maketitle
\section{Introduction}
\label{sec:introduction}

The remarkable progress in machine learning has inevitably trickled down and revolutionized the physical sciences~\cite{Carleo2019MLandPhysSciences}. In particular, machine learning has shown great promise for the numerical study of quantum many-body systems~\cite{Dawid2022}.  Machine learning and the quantum many-body problem share many similarities. Both fields are established in the intersection between applied linear algebra in large vector spaces, probability and statistics, and challenging optimization tasks. Therefore, it is only natural to wonder whether machine learning techniques can be applied to tackle the quantum many-body problem and further our knowledge of the underlying physical processes that govern the collective behaviour of interacting quantum particles. This review specifically addresses the application of machine learning techniques to existing numerical many-body methods to improve their accuracy and efficiency. 

The approximation of many-body wave functions by neural networks is one of the most promising avenues in the field~\cite{Carleo2017}. This parametric family is referred to as \textit{neural quantum states} (NQS). Neural networks provide a compact and accurate representation of the exponentially-scaling components of the many-body state. The accurate representation of quantum states is only possible thanks to the large expressive power of neural neural networks to represent complex functions~\cite{Cybenko1989Universality, HORNIK1989Universality, Gallant1988Universality}. It has been empirically and theoretically demonstrated that neural networks can approximate relevant quantum states whose entanglement entropy scales with the number of degrees of freedom of the problem~\cite{gao_efficient_2017, chen_equivalence_2018, sharir_neural_2022, Wu2023_2, gauvinndiaye2023mott} (volume-law scaling), while traditional variational states like the \textit{matrix product state}~\cite{White1992} (MPS) are not capable of efficiently describing highly-entangled states~\cite{Verstraete2004AreaMPS}. More generally, it has been shown that any efficiently-contractable \textit{tensor network} (TN) state can be efficiently mapped into a neural network~\cite{sharir_neural_2022, Wu2023_2}. 

The approximation of quantum states by neural networks relies on the gradient-based optimization of the weights and biases, since expectation values, overlaps and gradients thereof requires the use of Monte Carlo stochastic techniques for their estimation. It is this optimization where the greatest challenges for NQS arise, compared to other many-body methods like the \textit{density matrix renormalization group}~\cite{White1992} (DMRG) to optimize the parameters of a MPS. The characterization of the parameter optimization and its improvements remains an active field of research~\cite{Sinibaldi2023, Bukov2021learnin, Park2020GeometryofLearning, Westerhout2020Generalization, Luo_2023Training}.

Despite these challenges, NQS have demonstrated an enormous potential in the description of challenging many-body problems, including systems of interacting spins~\cite{Choo2019_J1-J2,fu_lattice_2022, Reh2023, Chen2023, fu_lattice_2022, astrakhantsev_broken-symmetry_2021} and systems of interacting fermions~\cite{nomura_restricted_2017, Luo2019backflow, robledo_moreno_fermionic_2022, pfau_ab_2020, Hermann2020Paulinet, Wu2023}.  These techniques have also demonstrated outstanding levels of accuracy in the description of the time evolution of many-body systems~\cite{Schmitt2020, Medvidovic2023, Gutierrez2022, Sinibaldi2023, Donatella2023}, the simulation of quantum circuits~\cite{Medvidovic2020} and quantum state tomography tasks~\cite{Torlai2018Tomography}. 

Other excellent review papers in the field are available, focusing on the practical implementation of these techniques~\cite{Carrasquilla2021review}, electronic structure problems~\cite{Hermann2023review} and neural-network architectures and applications~\cite{lange2024review}.

This review introduces, in a self-contained and pedagogical manner, the main concepts and theory behind the description of many-body wave functions using neural network approximations, their optimization and applications. The goal of this review is to provide the reader with the basis to understand the literature in the topic. Section~\ref{sec:vmc} introduces the general framework of variational Monte Carlo (Sec.~\ref{sec:variational_states}) and the need for Monte Carlo techniques to estimate expectation values of observables (Sec.~\ref{sec:localObservable}) and gradients with respect to variational parameters. Section~\ref{sec:vmc} also introduces the main sampling methods required to estimate such expectation values (Sec.~\ref{sec:sampling}) and four scenarios for variational parameter optimization: ground state search (Sec.~\ref{sec:optimization}), time evolution (Sec.~\ref{sec:tvmc}), overlap optimization (Sec.~\ref{sec:overlap_optimization}) and maximum likelihood estimation (Sec.~\ref{sec:mle}). These four techniques are the basis for the applications described in later sections. Section~\ref{sec:vmc} ends with a description of the fundamental concepts behind the imposition of symmetries in NQS calculations (Sec.~\ref{sec:symmetries}). Section~\ref{sec:applications} focuses on the application of the concepts introduced in  Section~\ref{sec:vmc} to relevant many-body systems, including the study of the ground state properties of systems of interacting spins (Sec.~\ref{sec:bosons_and_spins}), and systems of interacting fermions (Sec.~\ref{sec:chemistry}), as well as the unitary transformation of many-body wave functions to describe the time evolution of man-body systems and to simulate quantum circuits (Sec.~\ref{sec:unitary}). Sec.~\ref{sec:applications} concludes with a description of the application of NQS to the problem of quantum state tomography (Sec.~\ref{sec:tomography}). The scope of this manuscript is summarized in Fig.~\ref{fig:workflow}.

\section{Neural networks and Monte Carlo}
\label{sec:vmc}

\begin{figure*}
    \centering
    \includegraphics[width=1\textwidth]{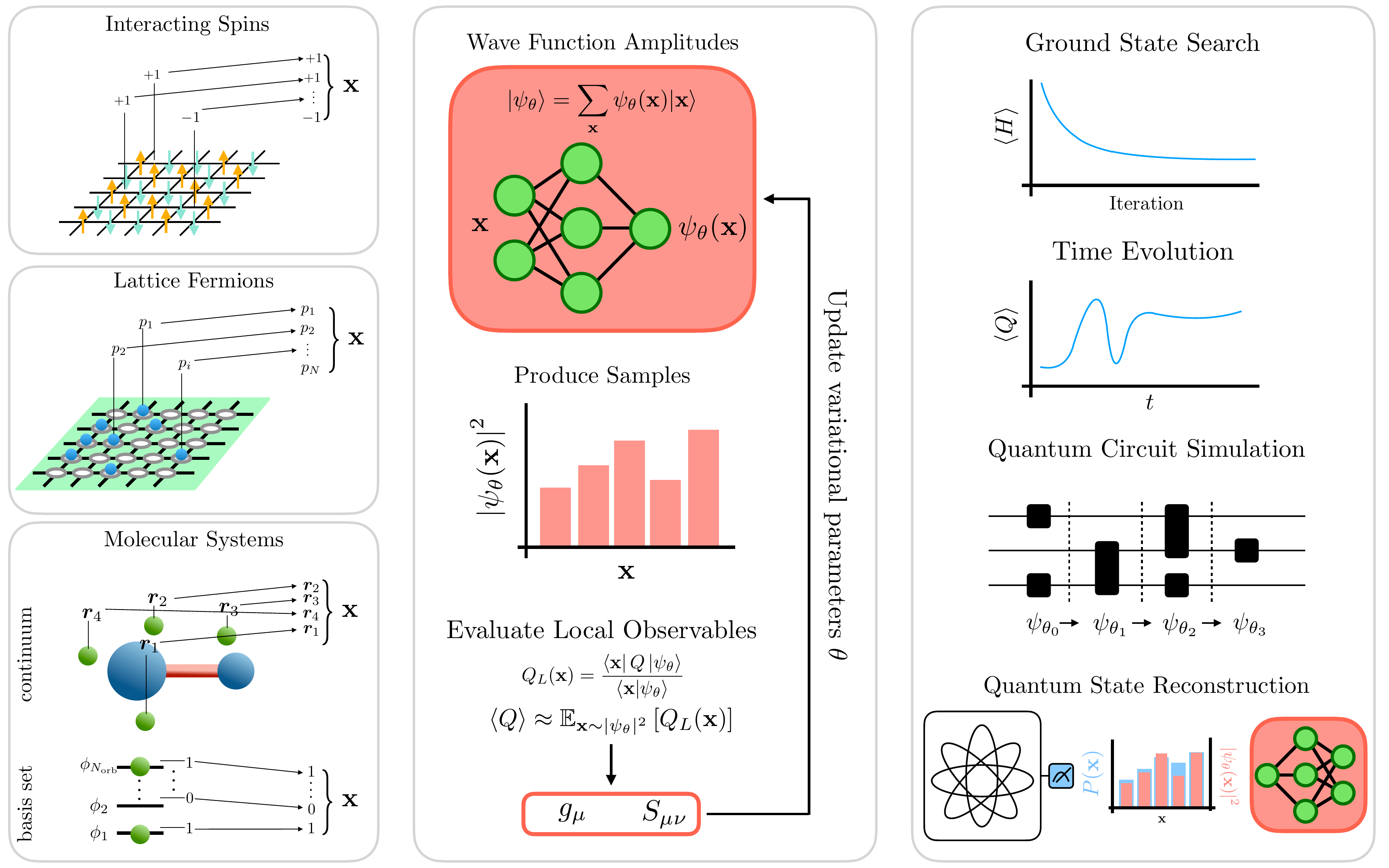}
    \caption{
        Neural Quantum States for many-body simulations. Different systems can be tackled by this technique, including quantum systems of interacting spins of qubits, fermionic lattice systems and interacting electrons in molecules, amongst others. The wave function amplitudes are given by the output of a neural network that takes as the input the corresponding basis element label $\xx$. Monte Carlo sampling is used to evaluate spectating values, overlaps and gradients, which are used to optimize the parameters of the neural networks. Different types of many body simulations can be realized, including the search of ground-state wave functions, the simulation of the time evolution of quantum states, the simulation of quantum circuits or quantum state reconstruction.
    }
    \label{fig:workflow}
\end{figure*}

In this section, we introduce the variational Monte Carlo (VMC) algorithm, motivating the optimization problem through the basic variational theorem of quantum mechanics (see Fig.~\ref{fig:workflow}). Stochastic estimators of Hilbert-space operators are introduced as a necessary approximation of full expectation values and Monte Carlo sampling algorithms are motivated as an essential tool in that process. We discuss both imaginary and real-time calculations through a common geometrical picture of Hilbert-space trajectories projected onto a variational manifold. Gradient-based optimization techniques of increasing complexity are introduced as a consequence of discretizing these trajectories. Finally, some common convergence metrics are proposed and discussed.

\subsection{Variational quantum state definition}
\label{sec:variational_states}

In order to follow through with the program laid out by the variational theorem, we consider a general Hilbert space $\mathcal{H}$, spanned by basis $\{\ket{\xx}\}$ of product states, or their completely symmetrized (or antisymmetrized) versions for systems of indistinguishable bosons (or fermions). The set of basis states satisfies the closure relation $\mathbbm{1} = \sum_{\xx} \ketbra{\xx}$. The following discussion applies to a wide variety of many-body systems. Some examples of systems and bases that they use to span the many-body Hilbert space include:

\begin{itemize}
    \item Spin systems: $\ket{\xx} \mapsto \ket{\uparrow \uparrow \hdots \uparrow}, \ket{\downarrow \uparrow \hdots \uparrow} \hdots$
    
    The basis is composed by all possible tensor products of single spin states in the reference quantization axis.

    \item Qubit systems: $\ket{\xx} \mapsto \ket{0 0 \hdots 0}, \ket{1 0 \hdots 0} \hdots$

    The basis is composed by all possible tensor products of single qubit states in the computational basis.

    \item Fermionic systems: $\ket{\xx} \mapsto \prod_{p = 1}^M \left(a^\dagger_p \right)^{n_p} \ket{\varnothing} \, .$

    The ket $\ket{\varnothing}$ refers to the Fock vacuum and $a^\dagger_p$ is the creation operator that adds a fermion to fermionic mode (or single-particle state or spin orbital) $p$. $M$ labels the number of single-particle states that span the Hilbert space. $n\in \{0, 1\}^M$ labels the occupancies of each fermionic mode. Many-body Hilbert space basis is obtained by the populating available fermionic modes in all possible ways. For a review of second quantization see Ref.~\cite{Negele1998Quantum}.

    \item Bosonic systems: $\ket{\xx} \mapsto \prod_{p = 1}^M \left(b^\dagger_p \right)^{n_p} \ket{\varnothing} \, .$
    
    The ket $\ket{\varnothing}$ refers to the Fock vacuum and $b^\dagger_p$ is the creation operator that adds a boson to bosonic mode (or single-particle state) $p$. $M$ labels the number of single-particle states that span the Hilbert space and $N$ the number of bosons in the system. $n\in \{0, N\}^M$ such that $\sum_{p = 1}^Mn_p = N$ labels the occupancies of each bosonic mode. Many-body Hilbert space basis is obtained by the populating available bosonic modes in all possible ways. For a review of second quantization see Ref.~\cite{Negele1998Quantum}.
\end{itemize}

Since the dimension of the Hilbert space for the above many-body systems grows exponentially or combinatorially with system size, a trial wave function based on the direct enumeration of all of the wave function amplitudes is not feasible beyond small system sizes. Instead, we introduce a general unnormalized variational quantum state
\begin{equation}
\label{eq:vqs}
    \ket{\psi _\theta} = \sum _{\xx} \psi _\theta (\xx) \ket{\xx} \; ,
\end{equation}
defined on the basis $\xx$ and parameterized by $P$ real or complex parameters $\theta $. Without loss of generality, we assume $\theta \in \mathbb{R} ^P $. For a discussion of all the possible cases, we refer the interested reader to an excellent overview in Ref.~\cite{Yuan2019}. The set $\mathcal{M} _\psi = \left\{ \ket{\psi _\theta} | \; \theta \in \mathbb{R} ^P \right\}$ is called the \textit{variational manifold} and, by definition, contains all states exactly representable by the trial state $\psi _\theta$. The goal of the techniques presented in following sections is to find the set of variational parameters $\theta$ such that $|\psi_\theta\rangle$ approximates a target many-body state as closely as possible. The variational state $\ket{\psi_\theta}$ is defined by its amplitudes, which are functions of the basis element labels $\xx$
\begin{equation}
    \psi_\theta: \xx \rightarrow \mathbb{C}.
\end{equation}

We require that the function $\psi_\theta$ can be evaluated at a polynomial cost in the number of degrees of freedom of the problem (number of spins, qubits, single-particle basis elements, particle number etc.). This requirement is the root of the approximate nature of the approach. The most general variational state $\psi_\theta$ is that represented by the lookup table of wave-function amplitudes for each basis state. However, due to the exponential or combinatorial growth of the Hilbert space dimension, this lookup table requires the storage of exponentially or combinatorially many amplitudes, making it not efficiently computable.

A neural quantum state is a variational wave function for which $\psi_\theta$ is partially or completely defined by a neural network whose weights and biases are part of the variational parameters $\theta$. A neural network defines a parametrized model family, whose functional form relies on the sequential composition of layers of parametrized transformations~\cite{Goodfellow2016MLBible}. They also satisfy a universal approximation theorem~\cite{Cybenko1989Universality, HORNIK1989Universality, Gallant1988Universality}, that states that for a sufficiently-large number of parameters, a neural network can approximate any target function to any specified accuracy. 

\subsection{Estimation of expectation values: the local observable trick}
\label{sec:localObservable}

The characterization of the physical properties of the many-body system (like the energy, susceptibilities...), as well as the parameter search for $\psi_\theta$ requires the efficient computation of expectation values of the form:
\begin{equation}
\label{eq:generic_expectation}
    \ev{Q} _{\psi _\theta} = \frac{\bra{\psi_\theta} Q \ket{\psi_\theta}}{\braket{\psi_\theta}} \; .
\end{equation}
where $Q$ is a generic many-body operator. The direct calculation of the numerator and denominator of the expectation value require the evaluation of a sum that contains a number of elements that grows exponentially/combinatorially with the number of degrees of freedom of the problem. Evaluating Hilbert space averages in Eq.~\ref{eq:generic_expectation} cannot be performed exactly. However, estimates can be made through Monte Carlo summation/integration. In general, expressions of the following form can be estimated\begin{equation}
\label{eq:monte_carlo_estimator}
    \ev{f(\xx)} _{\xx \sim p} = \frac{\sum _{\xx} p(\xx) f(\xx)}{\sum _\xx p(\xx)} \approx \frac{1}{N_\text{s}} \sum _{i = 1} ^{N_\text{s}} f(\xx _i) 
\end{equation}
using $N_\text{s}$ samples from the (unnormalized) distribution $p(\xx)$. This is a direct consequence of the Central Limit Theorem~\cite{Sorella2017QMCbible}. To make contact with the quantum VMC setting, we rewrite the energy expectation in Eq.~\ref{eq:generic_expectation} as:
\begin{equation}
\label{eq:local_operator}
    \ev{Q} _{\psi _\theta} = \frac{\sum_\xx \braket{\psi_\theta}{\xx} \bra{\xx} Q \ket{\psi_\theta}}{\sum_{\xx'}\braket{\psi_\theta}{\xx'} \braket{\xx'}{\psi_\theta}} = \sum _\xx \; p _\theta (\xx) \, Q_L (\xx )
\end{equation}
where
\begin{equation}
    p_\theta (\xx) \propto \left| \psi _\theta (\xx) \right| ^2 \quad \text{and} \quad Q_L (\xx) = \frac{\bra{\xx} Q \ket{\psi _\theta}}{\braket{\xx}{\psi _\theta}}
\end{equation}
defines the \textit{local observable} function $Q_L (\xx)$. Note that the evaluation of the local observable requires the calculation of the overlap $\bra{\xx} Q \ket{\psi_\theta}$. This is only possible if for every basis state $\ket{\xx'}$ the number of connected non-zero matrix elements to $\ket{\xx}$ grows at most polynomially with the number of degrees of freedom of the problem. The unbiased estimator to the expectation value of $Q$ is then obtained as the empirical average (Eq.~\ref{eq:monte_carlo_estimator}) of the associated local observable $Q_L$ evaluated at basis state configurations $\xx _i$ sampled from the state:
\begin{equation}
    \ev{Q} _{\psi _\theta} \approx \frac{1}{N_\text{s}} \sum_{i=1} ^{N_\text{s}} Q_L(\xx _i) = \ev{Q_L(\xx)} _{\xx \sim |\psi _\theta | ^2} \; .
\end{equation}

An operator of special relevance is the Hamiltonian of the system $H$. The associated local observable received the name of local energy $E_L = \flatfrac{\mel{\xx}{H}{\psi _\theta}}{\braket{\xx}{\psi_\theta}}$. An overview of relevant sampling-related algorithms and details is given in Sec.~\ref{sec:sampling}. In all cases, efficient calculation of many decorrelated samples is crucial for correctly resolving gradients and other averages.

\subsection{Sampling}
\label{sec:sampling}

As noted in Sec.~\ref{sec:optimization}, Monte Carlo sampling is used for evaluation of relevant averages in Eq.~\ref{eq:averages}. The goal is to obtain samples $\left\{ \xx _i : i = 1, \ldots , N_\text{s} \right\} \sim p(\xx)$ to evaluate sums of the type given in Eq.~\ref{eq:monte_carlo_estimator}, given an analytical probability distribution $p(\xx)$. Most commonly, well-established Markov-chain Monte Carlo (MCMC) black-box samplers from classical statistics literature are used~\cite{mcmc_handbook}. However, with certain restricted classes of trial states called \textit{autoregressive states}, efficient and exact sampling is possible. We review both cases in the following section.

\subsubsection{Markov Chain Monte Carlo sampling}

Often, MCMC algorithms are run in two steps -- first a new sample $\xx '$ is proposed from a current one $\xx $ and then it is either accepted or rejected according to the Metropolis-Hastings~\cite{Metropolis1953, Hastings1970} probability:
\begin{equation}
\label{eq:mh_accept}
    p_\text{accept} = \min \left(1, \flatfrac{p(\xx')}{p(\xx)} \right) \,
\end{equation}
assuming the usual constraints of detailed balance and ergodicity. It can be shown that biasing the random walk of proposed states with Eq.~\ref{eq:mh_accept} is guaranteed to converge to the correct distribution $p(\xx)$. We refer the the detail-oriented reader to Refs.~\cite{Newman1999, Strawderman2001}. In the context of wavefunction-baseed variational calculations such as VMC, the probability distribution is given by $p(\xx) \propto \left| \psi _\theta (\xx) \right| ^2$. Note that in the Metropolis-Hastings sampling procedure the probability distribution $p$ is not required to be normalized. Consequently, when using a Metropolis-Hastings sampler, the trial state does not need to be explicitly normalized.

In the context of VMC, the most popular choice of MCMC scheme is the simple \textit{local Metropolis} sampler, relying on small local changes to states $\xx$ to propose new ones $\xx '$. Depending on the physical system at hand, the simplest and most common update schemes can take different forms.

For systems with discrete degrees of freedom such as spins, we can simply flip a random spin in the state $\xx$ to obtain $\xx '$. This is the \textit{single-spin flip} update. If particle number or magnetization conservation needs to be enforced, pairs of up-down spins can be flipped simultaneously. We note that a sequence of such updates can be concatenated into a single larger update if acceptance rates are high enough.

For some systems, an efficient cluster update scheme can be constructed, greatly improving state space exploration. This is especially useful in the case of systems near criticality, where the correlation length is large and single-spin updates are inefficient. A popular cluster update scheme is the Wolff algorithm~\cite{Wolff1989} for the Ising model. For a review of cluster algorithms, we refer the reader to Ref.~\cite{Newman1999, Strawderman2001}.

In the case of continuous degrees of freedom such as electron positions $\xx$ in a molecular system, a small local update is usually defined as
\begin{equation}
    \xx ' = \xx + \delta \xx \; , \quad \text{where} \quad \delta \xx \sim \mathcal{N}(0, \Sigma ^2 ) \; .
\end{equation}

The normal distribution variance is set by $\Sigma ^2$, a hyperparameter tuned to achieve a certain acceptance rate. If the variance is too small, acceptance rates are high but individual updates stay too close to the original state, halting efficient exploration of the state space. If it is too large, acceptance rate drops because $\xx '$ can randomly hop too far outside of high probability regions and the sampler becomes inefficient. This version of the local Metropolis scheme is sometimes called the \textit{random walk Metropolis} (RWM) algorithm.

Another option for continuous systems is \textit{Hamiltonian Monte Carlo} (HMC) \cite{Neal2012, Hoffman2011, Carpenter2017}, offering a systematic way of making large steps in proposals while still keeping acceptance probabilities high, unlike RWM. This can result in lower MCMC autocorrelation times, allowing for treatments of larger systems with less overall runtime spent on sampling. For a probability distribution $p(\xx)$, HMC augments the configuration space with artificial momentum variables $\bm{\pi} = (\pi _1, \ldots, \pi _N) \sim \mathcal{N}(0, M)$:
\begin{equation}
\label{eq:hmc_hamiltonian}
    p(\xx) \propto \int \dd[N]{\bm{\pi}} \exp \left\lbrace -\frac{1}{2} \bm{\pi}^\top M^{-1} \bm{\pi} + \ln p(\xx) \right\rbrace
\end{equation}
for some positive-definite matrix $M$. The integrand in Eq.~\ref{eq:hmc_hamiltonian} can be read as an effective Boltzmann distribution $e^{-\beta \widetilde{H}}$ induced by an effective classical Hamiltonian $\beta \widetilde{H}(\xx, \bm{\pi})$. MCMC proposals can then be constructed through numerical integration of Hamilton's equations, starting from the current state $\xx$ -- particles in equilibrium following classical equations of motion conserve the desired Boltzmann distribution. Readers interested in an intuitive geometrical picture are referred to an excellent review in Ref.~\cite{Betancourt2017}.

\subsubsection{Autoregressive sampling}
\label{sec:autoregressive sampling}

Autoregressive sampling is an exact and \textit{embarrassingly parallel} algorithm for sampling probability distributions $p(\xx)$ of the special form
\begin{equation}
\label{eq:autoregressive}
\begin{gathered}
    p(\xx) = p(\xx_1,\ldots, \xx_N) = \prod _i p( \xx_i | \xx_{<i} ) = \\
    = p(\xx_1) \, p(\xx_2 | \xx_1) \, \cdots \, p(\xx_N | \xx_{N-1}, \ldots, \xx_2, \xx_1) \; ,
\end{gathered}
\end{equation}
where we have defined the set $\xx _{<i} = \left\{ \xx_1, \ldots , \xx_{i-1} \right\}$ by slight abuse of notation. If a distribution can be written in the form of Eq.~\ref{eq:autoregressive}, then a simple ancestral sampling approach can be used by iterating the following two steps for each $i=1,\ldots,N-1$:
\begin{enumerate}
    \item Sample $\xx_i \sim p(\xx_i | \xx_{<i} )$ and concatenate to $\xx_{<i}$.
    \item Use samples to define $ p(\xx_{i+1} | \xx_{<i+1} )$.
\end{enumerate}

We note that this procedure relies on being able to efficiently and exactly sample each conditional $p(\xx_i | \xx_{<i} )$. As a consequence, components $\xx_i$ have to be chosen to be sufficiently low-dimensional.

Some additional choices have to be made when generalizing autoregressive distributions to complex wavefunctions $\psi (\xx)$. For example, authors in Ref.~\cite{Barrett2022} choose to add a phase function globally by multiplying the expression in Eq.~\ref{eq:autoregressive} by $ e^{i \phi(\xx)}$ where $\phi$ is unrestricted while authors in Refs.~\cite{Hibat-Allah2020, Hibat-Allah2021}, choose to further decompose $\phi(\xx) = \sum_i \phi ( \xx_i | \xx_{<i} )$. Either way, the only restriction to including the phase is that $|\psi (\xx) | ^2$ has to have the structure of Eq.~\ref{eq:autoregressive}. 

\subsection{Stochastic optimization}
\label{sec:optimization}

The goal in variational calculations is to minimize the energy expectation
\begin{equation}
\label{eq:energy_expectation}
    E = E(\theta ) = \frac{\ev{H}{\psi _\theta}}{\braket{\psi _\theta}} \geq E_0 \; .
\end{equation}

The energy in Eq.~\ref{eq:energy_expectation} is now a function of parameters $\theta$ and an upper bound to the true ground state energy $E_0$ for all $\theta$. Our task then boils down to finding \textit{optimal} parameters $\theta _\text{opt} = \argmin E(\theta )$ that best approximate the ground state within $\mathcal{M} _\psi$. Modern gradient-based optimization methods are often used to find $\theta _\text{opt}$ in an iterative manner, starting from an initial guess. The expression for the relevant gradients $g = \nabla _\theta E(\theta )$ can be obtained by direct computation starting from Eq.~\ref{eq:energy_expectation}. However, in this review, we choose to derive a more general geometrical approach to this optimization problem from a physical point of view -- imaginary time evolution. The remainder of this section is devoted to connecting these ideas from statistical physics to such optimizers.

Imaginary time evolution well-known trick for systematically filtering out ground state $\ket{0}$ components out of arbitrary trial states $\ket{\psi}$, provided that the trial state is not orthogonal to the ground state:
\begin{equation}
    \ket{0} \propto \lim _{\tau \rightarrow \infty } e^{-\tau H} \ket{\psi} \, ,
\end{equation}
which can be made obvious by formally expanding $\ket{\psi} = \sum _n c_n \ket{n}$ in the energy eigenbasis $H\ket{n} = E_n \ket{n}$. We then have
\begin{equation}
\label{eq:imag_time_expansion}
\begin{gathered}
    e^{-\tau H} \ket{\psi} = \sum _n c_n \; e^{-\tau E_n} \ket{n} \propto \\
    \propto \ket{0} + \sum _{n > 0} \; \frac{c_n}{c_0} \; e^{-\tau (E_n - E_0)} \ket{n} \; ,
\end{gathered}
\end{equation}
assuming a non-degenerate ground state and $c_0 = \braket{0}{\psi} \neq 0$. The second term in Eq.~\ref{eq:imag_time_expansion} vanishes under the $\tau \rightarrow \infty$ limit because $E_n > E_0$ for all $n > 0$ by definition.

We apply imaginary time evolution to our trial state in small increments: $e^{-\tau H} = e^{-\delta \tau H} \cdots e^{-\delta \tau H} $. In other words, we iteratively update our parameters $\theta ' = \theta + \delta \theta$ so that:
\begin{equation}
\label{eq:param_update}
    \ket{\psi _{\theta + \delta \theta}} \propto e^{-\delta \tau H} \ket{\psi _\theta} \; ,
\end{equation}
where the proportionality sign accounts for the possible change in the overall normalization.

Expanding both sides in Eq.~\ref{eq:param_update} to first order in small parameter increments, we obtain $S \, \dot{\theta} = - g$ in the limit of $\delta \tau \rightarrow 0$, where $\dot{\theta} = \nicefrac{\dd \theta}{\dd \tau}$ and
\begin{equation}
    \label{eq:averages}
    \begin{gathered}
        S _{\mu \nu} = 2 \Re{ \ev{\OO ^\dagger _\mu \OO _\nu} - \ev{\OO ^\dagger _\mu} \ev{\OO _\nu \vphantom{\OO ^\dagger _\mu} } } \; ,\\
        g _\mu = 2 \Re{ \ev{\OO ^\dagger _\mu H} - \ev{\OO ^\dagger _\mu} \ev{H \vphantom{\OO ^\dagger _\mu} } } \; .
    \end{gathered}
\end{equation}

In Eq.~\ref{eq:averages}, $\mu, \nu, \ldots$ indices denote parameter dependence. Averages $\ev{\cdot} \equiv \nicefrac{\bra{\psi _\theta} \cdot \ket{\psi _\theta} }{ \braket{\psi _\theta} }$ are performed with respect to the trial state itself at imaginary time $\tau$. Diagonal operators $\OO _\mu$ are defined by $\partial _{\theta ^\mu} \ket{\psi _\theta} = \OO _\mu \ket{\psi _\theta}$ or
\begin{equation}
    \OO _\mu = \sum _{\xx} \frac{\partial  \ln \psi _\theta}{\partial \theta ^\mu} (\xx) \ketbra{\xx}{\xx} \; .
\end{equation}
A more detailed derivation of Eq.~\ref{eq:averages} starting from Eq.~\ref{eq:param_update} can be found in Appendix~\ref{appendix:vmc-eom}.

It must be noted that the above expression is only well defined if $\psi_\theta(\xx) \neq 0$. Through the Monte Carlo estimation of $S$ and $g$, one implicitly assumes that the support of the amplitudes and their gradients is the same. If this is not the case, the expression in Eq.~\ref{eq:averages} for $S$ and $g$ may contain a bias term~\cite{Sinibaldi2023}. We note that $g = \nabla _\theta E(\theta )$ is precisely the gradient that could have been obtained by direct differentiation of Eq.~\ref{eq:energy_expectation}. The matrix $S$ is commonly called the \textit{quantum geometric tensor} (QGT) or \textit{quantum Fisher information matrix}~\cite{Sorella1998, amari_natural_1998, stokes_quantum_2020} and acts like the metric tensor of the parameter manifold $\mathcal{M} _\psi$ induced by the distance in the surrounding Hilbert space $\mathcal{H}$ between un-normalized states defined in Eq.~\ref{eq:vqs}. In other words, it can be shown that the following expansion of the quantum fidelity $F(\cdot, \cdot)$ holds:
\begin{equation}
\label{eq:fidelity_metric}
\begin{gathered}
    F(\psi _{\theta + \delta \theta}, \psi _\theta) =
    \frac{\left\vert \braket{\psi _{\theta + \delta \theta}}{\psi _\theta} \right\vert ^2}{\braket{\psi _{\theta + \delta \theta}} \braket{\psi _{\theta}}} = \\
    = 1 - \frac{1}{2} \sum _{\mu \nu} S_{\mu \nu} \, \delta \theta ^\mu \, \delta \theta ^\nu + \cdots \; .
\end{gathered}
\end{equation}

\begin{figure}
    \centering
    \includegraphics[width=\columnwidth]{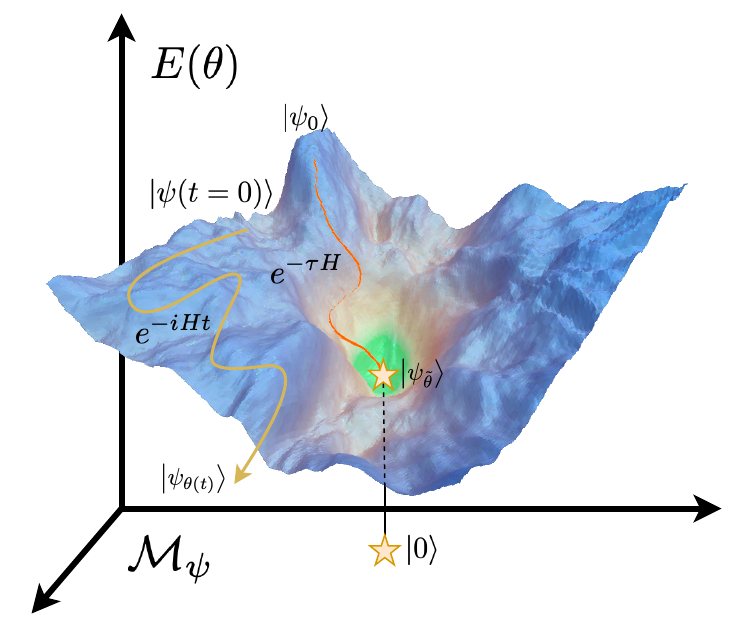}
    \caption{
        A schematic representation of optimization dynamics within the variational manifold $\mathcal{M} _\psi$. Optimization trajectories \textit{flow} towards lower energy states on the landscape $E(\theta)$ under imaginary time evolution operator $e^{-\tau H}$. In the case of real-time evolution $e^{-i H t}$, energy is conserved and trajectories are constrained to level-sets defined by the initial state $\ket{\psi (t = 0)}$.
    }
    \label{fig:landscape}
\end{figure}

The geometrical picture of this optimization procedure then becomes clearer -- imaginary time evolution induces a set of trajectories on the Hilbert space $\mathcal{H}$ connecting different initial states to to the ground state $\ket{0}$. Because we are limited to variational states $\ket{\psi _\theta}$, we project the true trajectory onto the variational manifold $\mathcal{M} _\psi$. The states on the projected path $\ket{\psi _{\theta (\tau)}}$ are parametrized by the solution to $S \, \dot{\theta} = - g$. The choice of how to discretize progress along this projection with sequential parameter updates is left to individual users. A schematic representation of such imaginary time dynamics is given in Fig.~\ref{fig:landscape}.

In addition, the $S$ matrix can be treated in several different approximations:

\begin{itemize}
    \item If no approximations are made and a simple Euler integrator is used to discretize the the top level ODE $S \, \dot{\theta} = - g$, one recovers the \textit{quantum natural gradient} (QNG) or the \textit{stochastic reconfiguration} (SR) update rule \cite{amari_natural_1998, Sorella1998}:
    \begin{equation}
    \label{eq:sr_update}
        \theta ' = \theta - \delta \tau \; S^{-1} g
    \end{equation}
    We note that more sophisticated integrators can be used as well, usually chosen from the Runge-Kutta (RK) family \cite{Butcher1963, Butcher2008}.
    \item If $S$ is instead approximated in block-diagonal form, the \textit{Kronecker-factored approximate curvature} (KFAC)~\cite{Martens2015} optimization scheme is recovered. Usually, $S _{\mu \nu}$ is assumed to vanish if $\mu$ and $\nu$ belong to different layers of the neural network state $\psi _\theta$ but other choices are possible.
    \item If we approximate $S \approx \mathbbm{1}$ in Eq.~\ref{eq:sr_update}, we obtain \textit{stochastic gradient descent} (SGD):
    \begin{equation}
        \theta ' = \theta - \eta \; \nabla _\theta E (\theta ) \, ,
    \end{equation}
    \noindent upon substituting $g_\mu = \partial _\mu E (\theta )$. We identify $\delta \tau$ with the learning rate $\eta$.
    \item Finally, if we keep $S = \mathbbm{1}$, we can use any of the popular optimizers from the deep learning literature such as Adam or Adamax~\cite{Kingma2014}, RMSProp~\cite{Graves2013} and others~\cite{Goodfellow2016MLBible}.
\end{itemize}

This hierarchy of increasingly complex optimizers is well-motivated by differential geometry on the manifold $\mathcal{M} _\psi$. Quantum natural gradient can be thought of as steepest descent optimization in curved space with metric tensor $S$, guaranteeing that the state described by the updated parameters is close in Fubini-Study metric to the initial state. For more insights into geometry on quantum variational manifolds, we refer the interested reader to an excellent review in Ref.~\cite{Hackl2020}.

While producing fast convergence and reliable results, the QNG/SR algorithm is sometimes avoided in practice due to the computational overhead for NQS with many parameters. The overhead is coming from the need to solve an $P \times P$ linear system at each step, where $P$ is the number of parameters. Recently, a group of authors in Ref.~\cite{Chen2023} introduced a computational trick allowing them to calculate the full preconditioned gradient $S^{-1} g$ using only a $N _\text{s} \times N _\text{s}$ matrix inversion, where $N _\text{s}$ is the number of samples. This simple trick reduces to exploiting a property of a Moore-Penrose pseudoinverse~\cite{Jacot2018} when evaluating $S^{-1} g$ in the finite-sample approximation
\begin{equation}
\label{eq:minsr}
    S^{-1} g =
    \left( \Bar{\OO} ^\dagger \Bar{\OO} \right)^{-1} \Bar{\OO} ^\dagger \epsilon =
    \Bar{\OO} ^\dagger \left( \Bar{\OO} \Bar{\OO} ^\dagger \right)^{-1} \epsilon
\end{equation}
where $\Bar{\OO} _{i \mu } = \bra{\xx _i} \left( \OO _\mu - \ev{\OO _\mu} \right) \ket{\psi _\theta} / \braket{\xx _i}{\psi _\theta}$ and $\epsilon _i = E_L (\xx _i)$. The new algorithm was named \textit{minSR} and opens the door to QNG on states with $\sim 10^6 - 10^7$ parameters.

In summary, the complete program reads:
\begin{enumerate}
    \item Choose a basis $\ket{\xx}$ and define a parameterized wavefunction $\psi _\theta (\xx)$.
    \item Approximate averages in Eq.~\ref{eq:averages} using MC samples and obtain $g = \nabla _\theta E$. (If optimizing using QNG, calculate the \textit{preconditioned gradient} $S ^{-1} g$ using an appropriate pseudoinverse $S^{-1}$.)
    \item Update parameters using a gradient-based optimizer.
    \item Go to 1. and repeat until convergence.
\end{enumerate}

After convergence has been reached, a measure of the quality of the obtained approximation is necessary. A common choice is the energy variance
\begin{equation}
    \Var E = \ev{H^2} _{\psi _\theta} - \ev{H} ^2 _{\psi _\theta}
\end{equation}
which vanishes as the true ground state is approached. However, recently, a version of the scaled variance, the \textit{V-score}, has been proposed by a wide group of authors in Ref.~\cite{Wu2023} as a more consistent measure of proximity to the ground state across different Hamiltonians and even Hilbert spaces. It is defined as:
\begin{equation}
\label{eq:v-score}
    \text{V-score} = \frac{N \; \Var E}{(E - E_\infty) ^2} \; ,
\end{equation}
for a system of $N$ degrees of freedom and an appropriately chosen reference energy $E_\infty$. In essence -- lower V-score is better. For a discussion of relevant details and V-scores across different methods and many-body Hamiltonians, we refer the reader to Ref.~\cite{Wu2023}.

\subsection{Time-dependent Variational Monte Carlo}
\label{sec:tvmc}

\begin{figure*}
    \centering
    \includegraphics[width=0.8\textwidth]{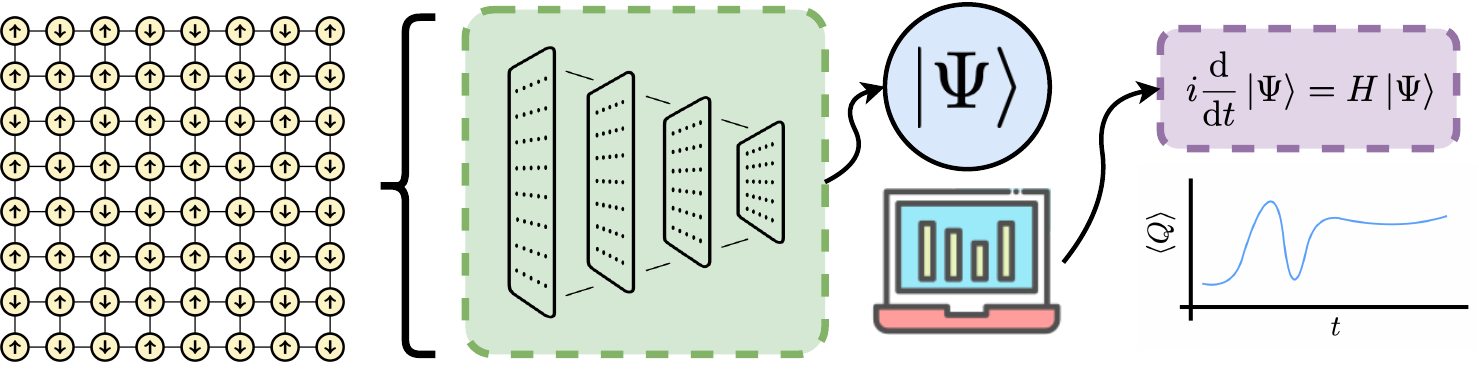}
    \caption{
        A schematic representation of the time-dependent variational Monte Carlo (t-VMC) algorithm. After defining the initial state as a variational wavefunction $\psi _{\theta(0)} (\xx)$, the exact time-evolved trajectory $e^{-i H t} \ket{\psi _{\theta(0)}}$ can be projected onto the variational manifold $\mathcal{M} _\psi$. The precise numerical implementation of this projection is an area of active research. However, Monte Carlo sampling of $\left| \psi \right| ^2$ is necessary in all cases, to estimate real-time dependence of a set of observables $Q$.
    }
    \label{fig:tvmc}
\end{figure*}

The geometrical picture introduced in section~\ref{sec:optimization} can be readily generalized to projections of other types of trajectories in the Hilbert space $\mathcal{H}$ onto the variational manifold $\mathcal{M} _\psi$. One of the most important developments in that field is unitary evolution in real time. The only difference is replacing $\tau \rightarrow i \, t$ in Eq.~\ref{eq:param_update}. Reusing notation from Eq.~\ref{eq:averages}, the relevant equation of motion then becomes $S \, \dot{\theta} = - f$, where
\begin{equation}
\label{eq:imag_grad}
    f _\mu = 2 \Im{ \ev{\OO ^\dagger _\mu H} - \ev{\OO ^\dagger _\mu} \ev{H \vphantom{\OO ^\dagger _\mu} } }
\end{equation}
and, expectedly, $\dot{\theta} = \nicefrac{\dd \theta}{\dd t}$. We note that the expression in Eq.~\ref{eq:imag_grad} is just the imaginary part of the energy gradient $g$ that was discarded in the case of imaginary time evolution in Eq.~\ref{eq:averages}. This method is called the time-dependent variational Monte Carlo (t-VMC). A good overview of the fundamentals can be found in Refs.~\cite{Sorella2017QMCbible, Yuan2019}.

For smaller systems, well-chosen variational states and shorter time horizons, Euler integration can be sufficient to produce good results. However, following the imaginary-time recipe and using a first-order method to step $t \rightarrow t + \delta t$ often leads to problems when scaling the algorithm. A part of the solution is to use higher-order and adaptive ODE integrators.

The reason is that the imaginary time evolution in Eq.~\ref{eq:imag_time_expansion} always has a well-defined limit while evolution in real time does not. In other words, $S \dot{\theta} = -g $ as an ODE is guaranteed to approach a fixed point for a wide range of initial conditions while its real time counterpart $S \dot{\theta} = -f$ is volume-preserving, up to projection errors. This is a direct consequence of the unitarity of the time-evolution operator $ \mathcal{U}(t) = e^{-i H t}$. As entanglement entropy grows under time evolution, the question becomes \textit{when}, not \textit{if} the simulation will become dominated by variational manifold projection errors.

In order delay this cutoff point and make t-VMC useful for larger systems at intermediate times, regularization methods are used when calculating $\dot{\theta} = -S^{-1} f$. Specifically, the $S$ matrix in Eq.~\ref{eq:averages} can become singular for trial states with many parameters like deep neural networks, preventing efficient inversion. Authors in Refs.~\cite{Carleo2012, Carleo2017_2, Ido2015} avoid this problem by restricted choices of trial states while authors in Ref.~\cite{Sinibaldi2023} choose to bypass explicit inversion by directly optimizing the overlap between the trial state at time $t$ and an estimate of the state at time $t + \delta t$. See Sec.~\ref{sec:unitary} for a discussion.

Refs.~\cite{Schmitt2020, Medvidovic2023} use spectral methods to define a custom pseudoinverse. The approach can be outlined as follows -- if $S$ is diagonalized as $S = U \Sigma U ^{\dagger}$ where $\Sigma = \text{diag}(\sigma ^2 _1, \ldots, \sigma ^2 _P)$ then one can replace
\begin{equation}
\label{eq:regularization}
    \Sigma ^{-1} _{\mu \nu} \rightarrow \frac{w(\sigma ^2 _\mu)}{\sigma ^2 _\mu} \delta _{\mu \nu} \quad \text{such that} \quad \lim _{\sigma ^2 \rightarrow 0} \frac{w(\sigma ^2)}{\sigma ^2} = 0 \; ,
\end{equation}
where $\delta _{\cdot \cdot}$ is the Kronecker delta. The function $w$ can be can be interpreted as \textit{spectral weight} and good choices usually strongly depend on different ansatz states.

In general, t-VMC regularization and introducing useful inductive biases in trial states is an open field of research.

\subsection{Overlap optimization}
\label{sec:overlap_optimization}

In some cases such as quantum circuit simulation~\cite{Medvidovic2020} real-simulation outside the traditional framework of t-VMC~\cite{Sinibaldi2023}, an explicit target state $\ket{\phi}$ is known. This should be contrasted with the traditional VMC setup in Sec.~\ref{sec:optimization} where the target state is implicit as a ground state of a given hamiltonian operator.

If one can easily evaluate the corresponding wave function $\phi (\vb{x})$, the following stochastic estimator of the quantum fidelity (introduced in Eq.~\ref{eq:fidelity_metric}) can be derived:
\begin{equation}
\label{eq:fidelity}
    F(\psi, \phi) = \frac{\left| \braket{\psi}{\phi} \right| ^2}{\braket{\psi} \braket{\phi}} = \ev{\frac{\phi (\vb{x})}{\psi (\vb{x})}} _{\psi} \ev{\frac{\psi (\vb{x})}{\phi (\vb{x})}} _{\phi} \; .
\end{equation}

The expression in Eq.~\ref{eq:fidelity} is manifestly independent of wave function renormalization. In practice, the imaginary part of the fidelity estimator in Eq.~\ref{eq:fidelity} is an artefact of finite sampling and is usually discarded.

If working with a parametrized state $\psi = \psi _\theta$ and a fixed target state $\phi$, then minimizing the \textit{infidelity} $\mathcal{D} (\psi, \phi) = 1 - F(\psi, \phi)$ requires corresponding gradients
\begin{equation}
\label{eq:infidelity_grad}
    \partial _\mu \mathcal{D}(\psi _\theta, \phi) = - F (\psi _\theta, \phi) \; \partial _\mu \ln F(\psi _\theta, \phi)
\end{equation}
with
\begin{equation}
\label{eq:fidelity_grads}
    \partial _\mu \ln F (\psi _\theta, \phi) = 2 \Re \left\{ \ev{ \OO ^\dagger _\mu} _{\psi _\theta} - \frac{ \ev{ \frac{\phi}{\psi _\theta} \OO ^\dagger _\mu} _{\psi _\theta}}{\ev{\frac{\phi}{\psi _\theta}} _{\psi _\theta}} \right\} \; .
\end{equation}

As an alternative to direct computation, we note that the expression for infidelity in Eq.~\ref{eq:fidelity} can be rewritten as an expectation value of an \textit{effective} Hamiltonian:
\begin{equation}
\label{eq:effective_hamiltonian}
    \mathcal{D}(\psi _\theta, \phi) = \frac{\ev{H ^\phi _\text{eff}}{\psi _\theta}}{\braket{\psi _\theta}}
    \; \; \text{with} \; \;
    H ^\phi _\text{eff} = \mathbbm{1} - \frac{\ketbra{\phi}}{\braket{\phi}} \; .
\end{equation}

Using the effective Hamiltonian from Eq.~\ref{eq:effective_hamiltonian} to compute gradients defined in Eq.~\ref{eq:averages} recovers the expression in Eq.~\ref{eq:fidelity_grads} and makes using the SR optimizer a natural choice in this case as well.

\subsection{Maximum likelihood estimation}
\label{sec:mle}

In contrast to the fidelity optimization case where the target state amplitudes $\phi(\xx)$ are known, we now focus on the task of finding an approximate representation $|\psi_\theta \rangle$ to a target state $|\phi\rangle$ without direct access to its amplitudes $\phi(x)$. Here we only have access to a training set of basis states $\mathcal{T} = \{ \xx^{[b]}$ \} on different bases $[b]$ sampled according to $\left|\bra{ \xx^{[b]}} \ket{ \phi } \right|^2$. Here the states $\ket*{\xx^{[b]}}$ are obtained as:
\begin{equation}
   \ket*{\xx^{[b]}} = U^{[b]} \ket{\xx},
\end{equation}
with $U^{[b]}$ the change of basis unitary that transforms $\{ \ket{\xx} \} \mapsto \left\{ \ket*{\xx^{[b]}} \right\}$. This is the typical problem encountered in quantum state tomography or quantum state reconstruction, where we have access to projective measurements on different  bases on a quantum experiment, and we seek to reconstruct the underlying quantum state. Within the umbrella of machine learning, this can be understood as a density estimation task. As such, the natural choice to find $\ket{\psi_\theta}$ is the optimization, with respect to $\theta$ of the KL divergence~\cite{Goodfellow2016MLBible} between the distributions defined by $\left|\bra*{ \xx^{[b]}} \ket{ \phi } \right|^2$ and $\left|\bra*{ \xx^{[b]}} \ket{ \psi_\theta } \right|^2$:
\begin{equation}
\label{eq:tomography_cost}
    \mathcal{D}_{\textrm{KL}} \left( |\phi |^2 \middle\| |\psi_\theta |^2 \right) = \sum_b \sum_{\xx^{[b]}} | \bar{\phi} ( \xx^{[b]} ) |^2 \ln \frac{| \bar{\phi} (\xx^{[b]} ) |^2}{| \bar{\psi}_\theta (\xx^{[b]} ) |^2} .
\end{equation}
$\ket*{\bar{\phi}}$ and $\ket*{\bar{\psi}_\theta}$ label the explicitly normalized counterparts to $\ket{\phi}$ and $\ket{\psi_\theta}$. In the above expression, the sum over $b$ runs over all (exponentially-many) basis choices, and the sum over all (exponentially-many) elements of each basis $\xx^{[b]}$. Different bases $b$ are required to be able to capture the phase structure of the target state. An unbiased estimator can be obtained as:
\begin{equation}
    \mathcal{D}_{\textrm{KL}} \left( |\phi |^2 \middle\| |\psi_\theta |^2 \right) \approx
    \frac{1}{|\mathcal{T}|} \sum_{\xx^{[b]} \in \mathcal{T}} \ln \frac{ |\bar{\phi} (\xx^{[b]} ) |^2}{| \bar{\psi}_\theta (\xx^{[b]} ) |^2}.
\end{equation}

The gradients of the empirical cost function are given by:
\begin{equation}
\begin{gathered}
    \partial_\theta \mathcal{D}_{\textrm{KL}} \left( |\phi |^2 \middle\| |\psi_\theta |^2 \right) \approx \\
    \frac{1}{N_\textrm{N}} \sum_{i=1} ^{N_\textrm{N}} \partial_\theta \ln |\psi_\theta (\xx _i)|^2
    - \frac{1}{|\mathcal{T}|} \sum_{\xx^{[b]} \in \mathcal{T}} \partial_\theta \ln | \psi_\theta (\xx^{[b]} ) |^2 .
\end{gathered}
\end{equation}

The first term, proportional to $1/N_\textrm{N}$, is the empirical average of the gradient of the normalization factor, evaluated over $N_\textrm{N}$ samples $\xx _i$ collected from $|\psi_\theta|^2$. The second term contains the gradient of the wave function amplitudes evaluated at basis states sampled according to the target state. Recall that:
\begin{equation}
    \psi_\theta (\xx^{[b]}) = \sum_{\xx} U^{[b]}_{\xx, \xx^{[b]}} \psi_\theta(\xx).
\end{equation}

We immediately recognize that in order for the KL optimization to be computationally efficient, the chosen bases that form the training set must satisfy that $U^{[b]}$ is $k-\textrm{local}$, so $\psi_\theta (\xx^{[b]} )$ can be efficiently evaluated. This technique was introduced in Ref.~\cite{Torlai2018Tomography}, adapted to a density-matrix formulation in Ref.~\cite{TorlaiMelko2018TographyDensityMatrix}, and applied to real experimental data in Ref.~\cite{Torlai2019TomographyRydberg}.

\subsection{Symmetries}
\label{sec:symmetries}

The imposition of discrete symmetries in NQS is an active field of research. Consider the case where the Hamiltonian of the system is invariant under the action of set of symmetry operations $\mathcal{S} = \{T\}$. Some examples of $T$ symmetry operations include translation, rotation or mirror symmetries or combinations thereof. The wave function amplitudes of the eigenstates of the Hamiltonian must satisfy the symmetry property:
\begin{equation}
\label{eq:symmetric amplitudes}
    \psi(T(\xx)) = \chi_T \psi(\xx),
\end{equation}
for all $T \in \mathcal{S}$. $\chi_T$ is the character of the symmetry operation $T$, and is a phase factor. The value of $\chi_T$ depends on the symmetry sector that $|\psi\rangle$ belongs to. 

The imposition of the symmetry requirements in Eq.~\ref{eq:symmetric amplitudes} in the parametric model $\psi^\text{S}_\theta$ can be achieved by the direct symmetrization of the trial state amplitudes of a non-symmetric ansatz~\cite{Nomura_2021}:
\begin{equation}
    \psi^\text{S}_\theta(\xx) = \sum_{T\in \mathcal{S}} \chi_T \; \psi_\theta\left(T^{-1}(\xx) \right).
\end{equation}

The direct symmetrization approach suffers from an increased computational cost of a factor of $|\mathcal{S}|$ every time a wave function amplitude or gradient needs to be evaluated. This computational overhead affects the estimation of expectation values as well as the sampling procedure. Alternatively, symmetry projections can be applied to a non-symmetric state after it has been optimized, obtaining improvements in the variational energy in the context of ground-state calculations~\cite{robledo_moreno_fermionic_2022}. The use of symmetry-equivariant neural networks like traditional convolutional neural networks~\cite{Choo2019_J1-J2, Reh2023, stokes_phases_2020}, their generalization to group-convolutions~\cite{roth_group_2021} or graph convolutional networks~\cite{fu_lattice_2022} also allow to bypass the explicit evaluation of $\psi_\theta(T(\xx))$ for all $T\in \mathcal{S}$ every time the variational state is symmetrized.

For ground state calculations, it is desirable to identify the symmetry sector that the ground state belongs to, to directly use the right characters for the symmetry operations. If this is not possible, separate ground state calculations must be carried projecting to all possible symmetry subspaces~\cite{Nomura_2021}. Amongst all optimized states on different subspaces, the ground state is identified with the one that has the lowest energy, as per the variational principle.

In the particular case where $\chi_T = 1$ for all $T \in \mathcal{S}$, the action of any non linear function $f$ on  $\sum_{T\in\mathcal{S}} \psi_\theta\left(T^{-1}(\xx)\right)$ yields a wave function ansatz that respects the symmetries of the Hamiltonian. Ref.~\cite{Reh2023} explores different functional choices for $f$ for systems of interacting spins. 

In some cases it is useful to fix the character \textit{a priori} to obtain excited states from the direct optimization of the expectation value of the Hamiltonian~\cite{choo_symmetries_2018}. If the target excited state is know to belong to a different symmetry sector to the ground state, the character of the symmetry can be fixed to the character corresponding to the symmetry subspace where the symmetries live. Consequently, the variational state is forced to live in a variational manifold that is not connected to the ground state.

It is also possible to impose non-Abelian anyonic symmetries in the context of NQS~\cite{vieijra_restricted_2020, Vieijra2021Jannes}. This is however requires a more involved technique.

It has been reported in numerous works that symmetry projections always improve the accuracy of the variational wave function. 

\section{Applications}
\label{sec:applications}

\subsection{Ground-state properties of spin systems}
\label{sec:bosons_and_spins}
\subsubsection{Hamiltonians}
Systems of interacting spins are the first class of many-body problems that were tackled by NQS~\cite{Carleo2017}. Systems of spins provide with a platform to study the fundamental properties of NQS and compare their expressive power against other variational methods, like tensor network methods. A general interacting spin Hamiltonian can be written as:
\begin{equation}
\label{eq:spin-hamiltonians}
    H = \sum_{ij} J_{ij} \left( \bm{\sigma}_i ^\top C \bm{\sigma}_j \right) - \sum_{i} \bm{\sigma}^\top _i  \bm{b}_i ,
\end{equation}
$\bm{\sigma}_i^\top = (\sigma^x_i, \sigma^y_i, \sigma^z_i)$ is the row-vector of Pauli operators acting on spin $i$. $J_{ij}$ represents the interaction strength between spins $i$ and $j$. $J_{ij}>0$ ($J_{ij}<0$) corresponds to antiferromagnetic (ferromagnetic interactions). $C$ is a $3 \times 3$ positive semidefinite matrix that characterizes the anisotropy of the interaction. $\bm{b}_i$ is a three-dimensional vector with the components of an external field applied to spin $i$. Many spin Hamiltonians can be written in this form, including the Ising, Heisenberg and $J_1-J_2$ models on different geometries. 

\subsubsection{Feedforward variational models}
One of the most widely used neural-network parametrizations for systems of spins is the Restricted Boltzmann Machine (RBM). The output of the RBM can be interpreted as the output of the Boltzmann distribution of a set of classical spins (input variables) that have Ising-like interactions with a collection of $H$ hidden spins, and optionally magnetic fields. Alternatively, the RBM can be seen as a two-layer perceptron with a very particular activation function. The strengths of the interactions and magnetic fields represent the weights $\bm{w}$ and biases $\bm{a}$ and $\bm{b}$ respectively of the neural network. Given the input row vector $\xx$, the RBM produces the output:
\begin{equation}
\label{eq:rbm}
    \psi_\theta(\xx) = e^{\xx \cdot \bm{a}} \prod_{i = 1}^H 2\cosh \left[ (\xx \cdot \bm{w} + \bm{b})_i \right].
\end{equation}
The number of hidden units $H$ controls the expressive power of the \textit{ansatz}. Real variational parameters $\bm{w}$, $\bm{a}$ and $\bm{b}$ make $\psi_\theta (\xx)$ to be real positive semidefinite, thus not being able to capture the intricate sign structures of many-body wave functions. The promotion of  $\bm{w}$, $\bm{a}$ and $\bm{b}$ to complex variables allows to capture the phases of the target state~\cite{Carleo2017}. The expressive power of the RBM can be further improved by allowing the input vector $\xx$ to contain both the sampled spin configuration along the quantization axis, as well as correlation functions (quadratic forms) of the spin configurations~\cite{Valenti2022CorrelationEnhanced}. This is known as the \textit{correlation enhanced RBM}. We also note that closely related feedforward models have been used for bosonic systems in Refs.~\cite{Saito2017, Saito2018, Pescia2022}.

Deeper and more involved neural network architectures, which in some cases include symmetry projections, convolutions and graph convolutions have been used to study challenging spin Hamiltonians like the $J_1-J_2$ model in square~\cite{Choo2019_J1-J2,fu_lattice_2022, Reh2023, Chen2023}, triangular, honeycomb and Kagome lattices~\cite{fu_lattice_2022}, or the Heisenberg model in the Pyrochlore lattice~\cite{astrakhantsev_broken-symmetry_2021}.

\subsubsection{Auto-regressive recurrent models}
\begin{figure}
    \centering
    \includegraphics[width=1\columnwidth]{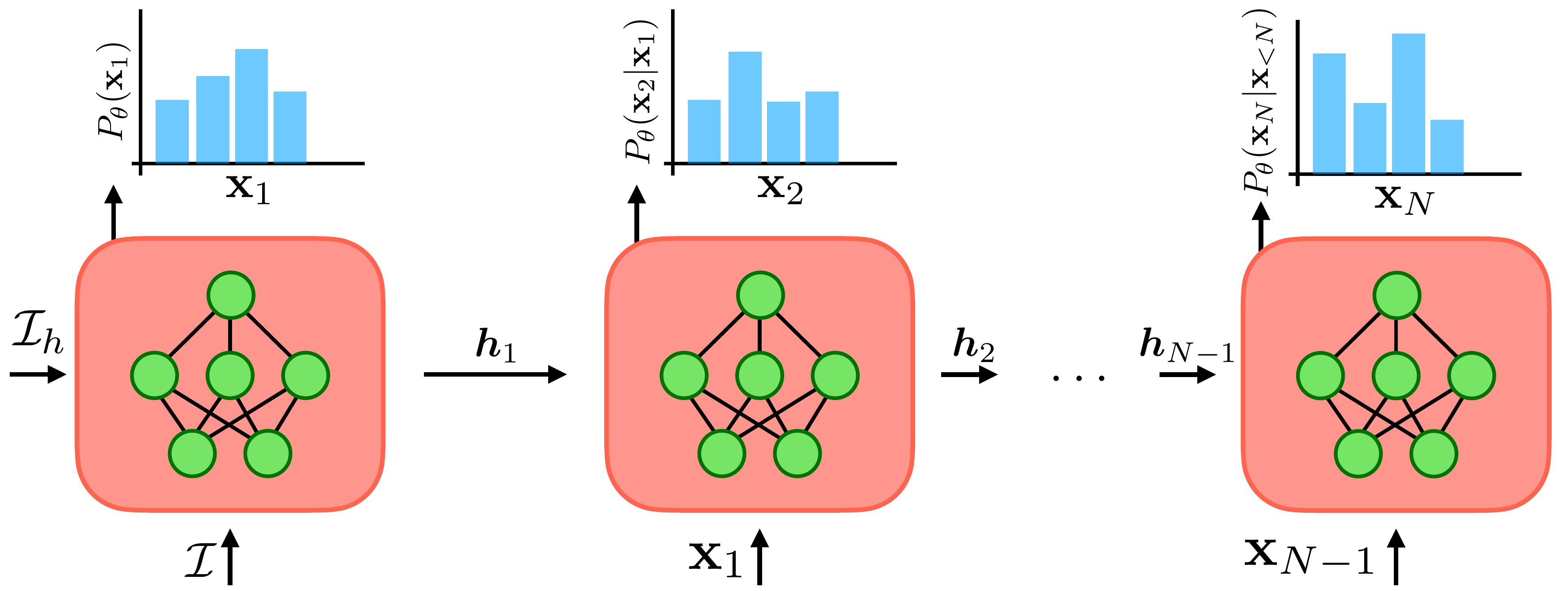}
    \caption{
        Schematic depiction of the use of a RNN to evaluate an auto-regressive model. A RNN is obtained by the repeated evaluation of the same feedforward block multiple times (all the blocks share the same parameters). Each feedforward block takes two sets of inputs and produces two sets of outputs. For the $i^\text{th}$ block the two sets of inputs consist on a \textit{hidden} state $\bm{h}_{i_1}$ and in our case the variable $\xx_{i-1}$. The two outputs produced by the block are $P_\theta(\xx_i | \xx_{<i})$ and the hidden state to be taken as an input by the following block $\bm{h}_i$. Note that the first block takes as inputs two fixed vectors $\mathcal{I}_h$ and $\mathcal{I}$. The information of the chain of conditional probabilities is encoded in the hidden states.
    }
    \label{fig:RNN}
\end{figure}

The field saw an important advancement with the introduction of recurrent neural networks (RNNs) to approximately represent the ground state of spin Hamiltonians~\cite{Hibat-Allah2020, roth2020_RNN_OG, Sharir2020_RNN_OG}. The RNN-based parametrization of a quantum state is defined by:
\begin{equation}
    \psi_\theta (\xx) = \sqrt{P_\theta(\xx)} \; e ^{i \phi _\theta (\xx)} ,
\end{equation}
where $\phi_\theta(\xx)$ is a parametric phase factor (which can be represented by a feedforward model) and $P_\theta(\xx)$ is an auto-regressive distribution model spanned by the chain rule of probabilities (see Eq.~\ref{eq:autoregressive}), where each conditional probability is normalized. Recurrent neural networks are a good candidate to parametrize the chain of conditionals needed to reconstruct $P_\theta$ as shown in Fig.~\ref{fig:RNN}. This class of models supports auto-regressive sampling and therefore is capable of producing perfect uncorrelated samples, as discussed in Sec.~\ref{sec:autoregressive sampling}. 

RNN architectures have successfully described physical properties of various spin models on different lattice geometries~\cite{Morawetz2021U-1, Hibat-Allah2020, Hibat-Allah2022RNN_symmetries_annealing, Hibat-Allah2023_topological_order, Czischek2022} as well as agreement with experimental observations of two-dimensional Rydberg atom array dynamics~\cite{Mendes2023-2}.

An open challenge for RNN-based parametrizations with auto-regressive sampling is the imposition of $U(1)$-like symmetries, such as magnetization of particle number conservation~\cite{lange2023Spectral, Morawetz2021U-1, Hibat-Allah2020, Hibat-Allah2022RNN_symmetries_annealing, barrett_autoregressive_2022, roth2020_RNN_OG, malyshev2023_RNN_U-1_symmetries}. 

\subsubsection{Expressive power, entanglement entropy and connections to tensor networks}
The understanding of the classes of states that can be efficiently\footnote{By efficiently we mean with an amount of computational resources that scales polynomially with the system size} represented by NQS is of imperative relevance to asses the utility of this class of trial states. 

In the space of feedforward architectures, it has been shown that the RBM can efficiently represents quantum states with volume-law entanglement entropy~\cite{deng_quantum_2017}. This result has been generalized to deep RNN and CNN architectures, showing that the deed architectures are capable of supporting volume-law scaling polynomially more efficiently than the RBM. These results suggest that deep architectures provide an advantage over shallow architectures to represent quantum states. Complexity-theory arguments support these statements~\cite{gao_efficient_2017}. Ref.~\cite{gao_efficient_2017} argues that most \textit{physical} quantum states can be represented by deep neural networks, while shallow architectures may not be able to efficiently represent some classes of physical quantum states. This means that while RBMs are capable of representing families of states with entanglement that follows a volume-law scaling, some of those states may not be relevant in the context of the simulation of physical systems.

By simply considering entanglement-entropy arguments, it might appear that NQS are more expressive that MPSs, since the MPS can only efficiently represent states whose entanglement follows an area-law~\cite{Verstraete2004AreaMPS}. This is however not the right evaluation in general, since there are physical quantum states that are not accessible to shallow NQS that can be represented by an MPS~\cite{gao_efficient_2017, Huang2021, Trigueros2023}. With the goal of finding the overlapping family of quantum states that can be both represented by TN states and NQS states, a body of work is centered around the search of exact mappings between TNs and NQS. Ref.~\cite{chen_equivalence_2018} provides an injective map between locally-interacting RBMs (most weights $\bm{w}$ set to zero) and an MPS. Ref.~\cite{chen_equivalence_2018} also show that some MPSs can be exactly mapped into a locally-interacting RBM. However, this work does not provide a recipe to exactly map an arbitrary MPS to a RBM.

The authors of Ref.~\cite{sharir_neural_2022} follow a different approach to construct mappings between TNs and NNs. An algorithm is provided to construct NN layers that efficiently perform tensor contractions. Therefore, a deep NN can be constructed to exactly represent an efficiently contractable TN, of course including the MPS. It was also found that the number of NN weights required for this mapping closely matches the contraction complexity of the TN. This results implies that deep NQS have strictly the same or more expressive power than efficiently contractable TNs. For example, any MPS can be exactly mapped into a deep NN whose total number of parameters grows polynomially with the bond dimension of the MPS, and whose depth grows logarithmically with the system size~\cite{sharir_neural_2022}. Given that deep NQS can represent states with entropy that follows a volume-law scaling and MPS cannot, this implies that MPS with polynomially-scaling bond dimension are s strict subset of deep NQS. Interestingly, these results can be generalized to RNNs~\cite{Wu2023_2}.

\paragraph{A word of caution.--} The results on the expressive power of NQS are really promising, and may even male the reader wonder why to use TN approaches like the DMRG at all if NQS provide a more expressive representation of quantum states. However, we would like to note that TN approaches like the DMRG pose two substantial advantages over NQS approaches. The first one is that some classes of TNs can be efficiently contracted, thus not requiring stochastic noisy estimates of expectation values. The second advantage is that some TN methods like the DMRG do not rely on gradient-based optimization techniques to optimize the parameters of the variational states. The reliance of NQS in gradient-based approaches make their accuracy strongly reliant on having a non-pathological variational energy landscape. However, at the time, the properties and topology of the  variational energy landscape for NQS parametrizations is not well understood. Further systematic investigations are required to characterize the optimization dynamics of NQS.

\subsection{Ground-state properties of fermionic systems}
\label{sec:chemistry}

\subsubsection{Hamiltonians}
Neural quantum states have been extensively employed to study systems of interacting electrons. The Hamiltonian for a system of $N$ interacting electrons in the presence of $N_\text{nuc}$ atomic nuclei, in the Born-Oppenheimer approximation form reads:
\begin{equation}
\label{eq:electrons continuum}
\begin{split}
    H & = -\frac{1}{2} \sum_{n} \vb{P}_n^2 + \sum_{n < m} \frac{1}{\left| \vb{r}_n - \vb{r}_m \right|} \\
    & - \sum_{n, A} \frac{Z_A}{\left| \vb{r}_n -\vb{R}_A \right|} + \sum_{A < B} \frac{Z_A Z_B}{\left| \vb{R}_A - \vb{R}_B \right|}.
\end{split}
\end{equation}

Note that atomic units have been used. In the equation above lowercase indices $n,\; m$ label electronic degrees of freedom while uppercase indices $A, \; B$ label nuclear degrees of freedom.  $\bm{P}_n$ and $\bm{r}_n$ are the linear momentum and position operators the electrons.  $\bm{R}_A$ labels the classical variable associated to the position of the atomic nuclei. NQS can be used directly in the continuum description of the problem~\cite{ruggeri_nonlinear_2018, pfau_ab_2020, spencer_better_2020, Hermann2020Paulinet, Entwistle2023, Glehn2022Psiformer, Casella_2D_electron_gas_2023, pescia2023electronGas, kim2023fermiGas, lou2023FermiGas}
. This is also known as the first quantization approach. 

It must also be noted that other fermionic many-body Hamiltonians have been addressed using neural quantum states. In particular, an area that is receiving increasing interest is the use of NQS to obtain the ground state properties of systems of interacting nucleons~\cite{Lovato_Hidden_nucleons_2022, gnech2023_hidden_nucleons, Fore_Hidden_nucleons_2022, Rigo2023_Slater_jastrow, Adams2021_Jastrow}.

The Hamiltonian in Eq.~\ref{eq:electrons continuum} is defined on a Hilbert space with an uncountable number of degrees of freedom. The same Hilbert space can be projected onto a truncated basis-set defined by a countable number $N_\textrm{orb}$ of orthonormal single-particle basis states or orbitals: $\left\{\ket{\phi_p}\; : \; p = 1, \hdots , N_\textrm{orb} \right\}$. The associated creation operators for the set of orbitals is defined by $\ket{\phi _p} = a ^\dagger _p \ket{\varnothing}$.

Since fermions live in the subspace of completely anti-symmetric states with $N$-particles, a suitable basis is obtained from explicitly anti-symmetrized products of single-particle basis states:
\begin{equation}
\begin{gathered}
    \ket{\phi_{p_1}\hdots \phi_{p_n} } = a^\dagger_{p_1} \hdots a^\dagger_{p_N} \ket{\varnothing} = \\
    = \frac{1}{\sqrt{N!}}\sum_{\Pi \in \textrm{S}_N} \xi(\Pi) \ket{\phi_{\Pi_1}} \otimes \hdots \otimes \ket{\phi_{\Pi_N}},
\end{gathered}
\end{equation}
where $\textrm{S}_N$ is the symmetric group of $N$ elements and $\Pi$ labels a permutations of the $\{p_1, \hdots, p_N \}$ indices whose sign is given by $\xi(\Pi)$. In this truncated representation of the many-body Hilbert space the Hamiltonian in~\ref{eq:electrons continuum} can be written in terms of creation and annihilation operators:
\begin{equation}
\label{eq:electrons discrete}
\begin{gathered}
    H = \sum_{\substack{pq \\ \sigma}} h_{pq}  c^\dagger_{p\sigma} c_{q\sigma}
    + \sum_{\substack{pqrs \\ \sigma \tau}} \frac{h_{pqrs}}{2} c^\dagger_{p\sigma} c^\dagger_{q\tau} c_{s \tau} c_{r \sigma} 
\end{gathered}
\end{equation}
where $\sigma, \tau \in \{\uparrow, \downarrow\}$ and $h_{pq}$ and $h_{pqrs}$ are the one- and two-body matrix elements. Their values depends on the choice of single-particle orbitals and can be extracted from quantum chemistry packages~\cite{Pyscf1, pyscf2, pyscf3}. The interacting-electron Hamiltonian is number and magnetization preserving as well as $SU(2)$ symmetric. Note that the indices $p, q, r, s$ run from $1$ to $N_\textrm{orb}$. Consequently, the number of terms in the Hamiltonian scales as $\mathcal{O} \left((N_\textrm{orb})^4 \right)$. The asymptotic scaling on the number of terms in the Hamiltonian is multiplied to the asymptotic scaling of the wave function evaluation. Therefore, the direct study of the interacting electron Hamiltonian in second quantization becomes computationally expensive rapidly. NQS have also been extensively used to study systems of interacting electrons on a discrete basis set~\cite{xia2018_OG-NQS-QChem, choo_fermionic_2020, yang2020_NQSQChem, Yoshioka2021solids, rath_framework_2023, sureshbabu2021_NQSQC,zhao_2022autoregressiveNQS, barrett_autoregressive_2022, inui_determinant-free_2021, Nys2022variational, nomura_restricted_2017, stokes_phases_2020, Luo2019backflow, robledo_moreno_fermionic_2022, RobledoMoreno2023basisrotations, gauvinndiaye2023mott, lange2023Spectral}.

Hamiltonians in Eqs.~\ref{eq:electrons continuum} and~\ref{eq:electrons discrete} are referred to as \textit{ab-initio} Hamiltonians. It is at this point where physically-motivated approximations can be implemented to reduce the number of terms in the two-body part of the Hamiltonian. Under reasonable assumptions of locality of tightly bound electrons in periodic arrangements of atoms led Hubbard~\cite{Hubbard1963} to derive a simplified interacting electron Hamiltonian. The simplest version of the Hubbard Hamiltonian considers only a single orbital per atom (single-band), one-body processes between adjacent atoms and interaction processes only between electrons in the same atom, yielding the Hamiltonian
\begin{equation}
\label{eq:fermi-hubbard}
    H = - t \sum_{\ev{p, q}, \sigma} \left( c^\dagger_{p\sigma} c_{q\sigma}  + c^\dagger _{q\sigma} c_{p\sigma} \right) + U \sum_p n_{p\uparrow} n_{p\downarrow},
\end{equation}
where $t$ is the hopping amplitude and $U$ is the onsite interaction strength. The Hubbard model provides a minimal description of locally interacting electrons hopping between adjacent sites of a given lattice. Despite its simple appearance, the complete understanding of its ground-state properties is still a very active field of research~\cite{QinHubbardSiperconductivity, baldelli2023fragmented, Wietek2021Hubbard_trialgular, Zheng2017stripes, Lanata2017FullyConnectedHubbard}. 

\subsubsection{Symmetrization considerations}

Fermionic systems pose a unique challenge due to their particle statistics. As a consequence, the creation and annihilation operators satisfy the commutation relations:
\begin{equation}
    \label{eq:anti-commutators}
    \left\{ a_p, a_q \right\} = \left\{ a^\dagger_p, a^\dagger_q \right\} = 0
    \, ; \quad \text{and} \quad
    \left\{ a_p, a^\dagger_q \right\} = \delta_{pq} \, .
\end{equation}

A consequence of the above commutation relations is that $a_p^2 = \left(a^\dagger_p\right)^2 = 0$ (Pauli's exclusion principle).

The basis of the subspace of completely anti-symmetric states can be spanned in terms of either canonically-ordered or non-canonically-states. This has an impact on the symmetry properties of the wave function amplitudes $\psi_\theta(\xx)$. 

First consider the case where the Hilbert space is spanned in terms of non-canonically-ordered states:
\begin{equation}
    \{ \ket{\phi_{p_1} \hdots \phi_{p_N}} \: : \: p_i = 1, \hdots, M\}.
\end{equation}
From now on the symbol $\xx$ represents the index set $p_1, \hdots, p_N$ and $\ket{\xx} := \ket{\phi_{p_1} \hdots \phi_{p_N}}$. The basis expansion in terms of non-canonically-ordered states is highly redundant since:
\begin{equation}
\label{eq:anti-symmetry basis states}
    \ket{\xx} = \xi(\Pi) \ket{\Pi(\xx)}
\end{equation}
for all $\Pi \in \text{S}_N$, and $|\text{S}_N| = N!$. This is a consequence of the anti-commutation relations of the creation operators. This is reflected in the $1/N!$ prefactor in the resolution of the identity:
\begin{equation}
    \mathbbm{1} = \frac{1}{N!} \sum_\xx \ketbra{\xx}.
\end{equation}
A direct consequence of Eq.~\ref{eq:anti-symmetry basis states} is that the wave function amplitudes must satisfy:
\begin{equation}
\label{eq:anti-symmetric amplitudes}
    \psi_\theta(\xx) = \xi(\Pi) \psi_\theta( \Pi (\xx ))
\end{equation}
for all $\Pi \in \text{S}_N$. Wave function amplitudes are anti-symmetric functions of the orbital labels. 

Alternatively, the subspace of completely anti-symmetric states can be spanned in terms of canonically-ordered states $\{ \ket{\encircled{\xx}} \}$, where we choose the ordering where the orbitals are populated in a defined order: $\{\encircled{\xx} = p_1, \hdots, p_N \; : \; p_1 < \hdots < p_N\}$. Consequently, the resolution of the identity reads: $\mathbbm{1} = \sum_{\encircled{\xx}} \ket{\encircled{\xx}} \bra{\encircled{\xx}}$, and the variational wave function is written as:
\begin{equation}
 \ket{\psi_\theta} = \sum_{\encircled{\xx}} \psi_\theta(\encircled{\xx}) \ket{\encircled{\xx}},
\end{equation}
where the amplitudes are related to those of the non-canonically-ordered expansion by $\psi_\theta(\encircled{\xx}) = \xi \left(\Pi_{\xx \rightarrow \encircled{\xx}}\right) \psi_\theta(\xx)$. $\Pi_{\xx \rightarrow \encircled{\xx}}$ labels the permutation that brings an arbitrary electronic configuration $\xx$ to its ordered counterpart $\encircled{\xx}$. 
 
Since there is a on-to-one map between canonically-ordered configurations $\encircled{\xx}$ and the occupancy of the orbitals $n$, variational wave functions spanned in terms of canonically-ordered states are commonly taken to be functions of the orbital occupancies $\psi_\theta: n \mapsto \mathbb{C}$. 

In the canonically-ordered representation, every-time that an operator $Q$ (which is to be written in terms of creation and annihilation operator) acts on a basis state, the resulting states must be brought back to their canonically-ordered form, which is accomplished by computing the sign of the permutation of the transformations. The book-keeping of signs of permutations can be conceptualized by the Jordan-Wigner mapping~\cite{JordanWigner1928}. However, in practice it is more efficient to keep track of the signs by hand. 

The variational states used to describe systems of interacting fermions can be divided into two groups: non-determinant- and determinant-based parametrizations. Traditionally, non-determinant-based parametrizations are employed when the Hilbert space is spanned in terms of canonically-ordered states~\cite{xia2018_OG-NQS-QChem, choo_fermionic_2020, yang2020_NQSQChem, sureshbabu2021_NQSQC, Yoshioka2021solids, rath_framework_2023, barrett_autoregressive_2022, zhao_2022autoregressiveNQS, lange2023Spectral}, while determinant-based parametrizations are employed when the Hilbert space is spanned by non-canonically-order states~\cite{nomura_restricted_2017, stokes_phases_2020, Rigo2023_Slater_jastrow,  Adams2021_Jastrow, robledo_moreno_fermionic_2022, pfau_ab_2020, Hermann2020Paulinet, Luo2019backflow}. There are however exceptions~\cite{inui_determinant-free_2021, RobledoMoreno2023basisrotations}. 

\subsubsection{Non-determinant-based parametrizations}

Non-determinant, neural-network-based parametrizations were first introduced to tackle the electronic structure properties of small molecules described by minimal basis sets~\cite{xia2018_OG-NQS-QChem, choo_fermionic_2020, yang2020_NQSQChem, sureshbabu2021_NQSQC}.  Following the tradition in the field, most of this works employed the RBM as the trial wave function. The RBM \textit{ansatz} has also been employed to study the ground-state properties of solids, including hydrogen chains and three-dimensional $\text{LiH}$ in their minimal basis sets, as well as graphene described by a correlation-consistent basis set~\cite{Yoshioka2021solids}. The work in Ref.~\cite{Yoshioka2021solids} also included Krylov susbpace methods to accurately reconstruct the spectra of excitations of these systems. Other classes of machine learning models based on Gaussian processes have also been proven successful in tackling molecular systems on a basis set~\cite{rath_framework_2023}. 

RNNs have also been used to describe the ground-state properties of the electronic degrees of freedom of molecules~\cite{barrett_autoregressive_2022, zhao_2022autoregressiveNQS}, as well as lattice models~\cite{lange2023Spectral}. Ref.~\cite{lange2023Spectral} also provides a path to computing excitations using NQS-based parameterizations. The main advantage behind the use of RNNs as opposed to feedforward neural networks is that, thanks to the auto-regressive sampling support, they produce uncorrelated samples. This may be of great relevance in the description of molecular systems projected on a basis set, since there are some basis choices where the wave function is concentrated around a small number of electronic configurations, while the amplitude of the remaining configurations decays exponentially. Thus, posing a challenge for Markov-Chain-based samplers. 

Lattice models, like the Hubbard model have also been addressed by deep NQS~\cite{inui_determinant-free_2021}. Alternative fermion-to-spin mappings can be used in this context~\cite{Nys2022variational}.

\subsubsection{Determinant-based parametrizations.}
\paragraph{Slater determinant.-- } The determinant-based parametrizations use Slater determinant states as a starting point, and add correlations to them. The Slater determinant state $\ket{\psi_\text{SD}}$ for a system with $N$ particles is obtained by populating $N$ variational orbitals:
\begin{equation}
    \ket{\psi_\text{SD}} = \varphi^\dagger_1 \hdots \varphi^\dagger_N \ket{\varnothing}, 
\end{equation}
where $\varphi^\dagger_\alpha = \sum_{p} \Phi_{p,\alpha} a^\dagger_p$. The rectangular matrix $\Phi \in \mathbb{C}^{M \times N}$ controls the definition of the variational orbitals, and its matrix elements are the variational parameters. The wave function amplitudes of the Slater determinant are given by:
\begin{equation}
\label{eq:Slater determinant}
    \psi_\text{SD} (\xx) = \bra{\xx} \ket{\psi_\text{SD}} = \det 
    \begin{bmatrix}
        \Phi_{p_1, 1} & \hdots & \Phi_{p_1, N} \\
        \vdots & \ddots & \vdots \\
        \Phi_{p_N, 1} & \hdots & \Phi_{p_N, N} \\
    \end{bmatrix}. 
\end{equation}
Note that the exchange of two of the indices in $\xx$ swaps two rows in the determinant, thus making the amplitudes anti-symmetric, as required by Eq.~\ref{eq:anti-symmetric amplitudes}. Also note that, the same variational state can be used even when the Hilbert space is spanned in terms of canonically-ordered states. In this case, one just had to replace $\xx$ by $\encircled{\xx}$ in Eq.~\ref{eq:Slater determinant}. When working in the continuum the Slater determinant is obtained by replacing the columns in $\Phi$ (which can be understood as a lookup table of single-particle orbital amplitudes) by functions that depend on the coordinate of one particle $\Phi_\alpha (\xx_i)$. The Slater determinant is the anti-symmetrization of a product state and therefore is incapable of capturing electronic correlations.

\paragraph{Jastrow-like correlation factors.-- }  Correlations can be achieved by the multiplication of a permutation-symmetric correlation factor $\mathcal{S}$ (Jastrow factor~\cite{Jastrow1955}):
\begin{equation}
    \psi_\theta(\xx) = \psi_\text{SD}(\xx) \mathcal{S}_\theta (\xx)
\end{equation}
The Jastrow factor can be parametrized by a neural network. When working on a basis-set, the Jastrow factor is typically taken to be a function of the fermionic mode occupancies: $\mathcal{S} :n \mapsto \mathbb{C}$, thus becoming automatically symmetric. The use of neural-network correlation factors was first introduced in Ref.~\cite{nomura_restricted_2017}, where a RBM with real parameters was used as a correlation factor on top of a pairing wave function, focusing on the ground-state properties of the Hubbard model in square geometries. The RBM of real parameters can only produce positive semidefinite outputs. Consequently, the Jastrow factor in Ref.~\cite{nomura_restricted_2017} can only correct the amplitudes of the reference anti-symmetric state. The generalization of this idea to more complex neural-network architectures was introduced in Ref.~\cite{stokes_phases_2020}, where $\mathcal{S}$ is parametrized by the product of two CNN with skip connections. The first one, producing a positive semidefinite output is in charge of correcting the amplitudes of the reference Slater determinant, while the second one produces an output in the interval $(-1, 1)$, in charge of altering the nodal structure of the reference state. This \textit{ansatz} is universal in the description of wave functions of interacting systems of fermions described by a countable number of degrees of freedom~\footnote{In the context of NQS for fermionic systems, universal representation claims are based on lookup-table arguments, thus not providing any practical information on the classes of states that can be represented by these \textit{ansatze}.}. Ref.~\cite{stokes_phases_2020} focuses on the study of the ground-state phase diagram of spin-polarized lattice fermions. Uncorrelated sampling can also be achieved with Slater-Jastrow functional forms, where the Jastrow factors are given by RNNs~\cite{Humeniuk2023_RNN_Slater}. Neural network Jastrow factors have also been extensively used to study systems of interacting nucleons~\cite{Rigo2023_Slater_jastrow,  Adams2021_Jastrow}.

\paragraph{Backflow correlations.--} Alternatively, correlations can be incorporated in Slater determinants through Backflow transformations~\cite{Sorella2017QMCbible, Wigner1934Backflow, Feynman1954Backflow, Feynman1956Backflow, Kwon1993Backflow, Kwon1998Backflow}. In the continuum representation of the problem this involves replacing the single-particle coordinates that are the inputs to the orbitals that enter the determinant: $\Phi_\alpha(\xx_i) \rightarrow \Phi_\alpha(\bar{\xx}_i)$, where $\bar{\xx}$ is referred to as backflow coordinates:
\begin{equation}
    B_\theta : \xx \mapsto \bar{\xx},
\end{equation}
where $B_\theta$ is a parametric function that maps the electron coordinates to the backflow coordinates. For the determinant to maintain its ant-symmetry properties, the function $B_\theta$ is required to be permutation-equivariant. Ref.~\cite{ruggeri_nonlinear_2018} represented the backflow transformation using neural-network-like functional forms. Non-linear and parametrized backflow transformations are composed and applied to the input electron coordinates. This goes by the name of iterative backflow. Later, Refs.~\cite{pfau_ab_2020, spencer_better_2020}, introduced the \textit{FermiNet} neural network architecture to parametrize $B_\theta$. The FermiNet architecture is agnostic to known properties that the wave function must satisfy, like the Kato cusp conditions~\cite{Kato1957Cusps}, and is expected to learn to reproduce the correct behaviour. It has been empirically demonstrated that the FermiNet \textit{ansatz} can achieve great levels of accuracy in the description of the ground state properties of molecular systems.  The \textit{PauliNet}~\cite{Hermann2020Paulinet} wave function ansatz is also based on the parametrization of the backflow transformation using neural networks, but exactly incorporates the Kato cusp conditions. The PauliNet \textit{ansatz} also considers a Jastrow-like factor to account for the electron-electron cusp conditions. Like the FermiNet \textit{ansatz}, the PauliNet variational state achieves great levels of accuracy in molecular systems. The PauliNet \textit{ansatz} has also been proven successful to obtain low-laying exited states in molecules~\cite{Entwistle2023}, by optimizing the expectation value of the Hamiltonian with an overlap penalty to the ground states. More recently, a novel approach exploits the properties of determinants to ensure linear independence of a number of excited states~\cite{pfau2023excited_determinant}, allowing for the simultaneous search of a number of excited states above the ground state. Motivated by the recent success of large language models, which strongly rely on attention mechanisms, self-attention parametrizations have been employed to parametrize the backflow transformation $B_\theta$~\cite{Glehn2022Psiformer}. Since self attention mechanisms in essence represent a parametrized two-point correlation function, the idea of using self-attention mechanisms to parametrize wave functions can be seen as the generalization of the correlation-enhanced NQS introduced in Ref.~\cite{Valenti2022CorrelationEnhanced}.

All of the previously mentioned studies focus on molecular systems. Recently, the neural-network parametrization of backflow transformations has been applied to periodic systems, including the two-dimensional electron gas~\cite{Casella_2D_electron_gas_2023}, the three-dimensional electron gas~\cite{pescia2023electronGas}, the three-dimensional Fermi gas~\cite{kim2023fermiGas, lou2023FermiGas} and Moire systems~\cite{luo2023Moire}.

The backflow parametrizations of fermionic wave functions constitute a universal wave function approximator for continuous degrees of freedom~\cite{pfau_ab_2020}.

When working on a basis set, the backflow wave function is obtained by allowing the matrix elements of $\Phi$ to be configuration-dependent~\cite{Tocchio2008Backflow, Tocchio2011Backflow}:
\begin{equation}
    \Phi_\text{BF} : \xx \mapsto \mathbb{C}^{M\times N},
\end{equation}
where $\Phi_{\text{BF}}$ is required to be symmetric with respect to permutations in $\xx$, often-times being a function of the fermionic mode occupancies. The use of neural networks to parametrize $\Phi_\text{BF}$ to study lattice models was introduced in Ref.~\cite{Luo2019backflow}, achieving great levels of accuracy in the ground state properties of the Hubbard model on a range of geometries. Backflow wave functions are also a universal wave function approximator for fermionic systems with a countable number of degrees of freedom~\cite{Luo2019backflow}.

\paragraph{Neural-network constrained hidden fermion determinant states.--}
Traditional hidden particle approaches have allowed to obtain significant insight on the properties of fermionic lattice models like the Hubbard model and its variants like the $t-J$ model~\cite{barnes_1976,barnes_1977, Kotliar1986SlaveBosons,kotliar_largeN, li_rotinv_1989, fresard_1992, slavefermion1, slavefermion2, Florens2002SlaveRotor, Florens2004Slaverotor, Medici2005slavespins,lechermann_2007, lanata_ghostGA_2017, frank_ghost_2021, guerci_2019, guerci_phd_2019, ancilla_zhang_2020, ancilla_nikolaenko_2021, coleman_1984}. Hidden particle approaches are rooted on the augmentation of the Hilbert space of the problem by the addition of extra degrees of freedom, which can be of spin, bosonic or fermionic nature. In the augmented Hilbert space, a mean-field theory can be considered. A subsequent projection to a subspace of the augmented Hilbert space identified with the physical degrees of freedom allows to recover a description of the correlations in the system. In the traditional formulation, the characteristics of the added degrees of freedom and the projection are chosen based on the expected properties that one tries to describe, thus posing an inductive bias.

\begin{figure*}
    \centering
    \includegraphics[width=.7\textwidth]{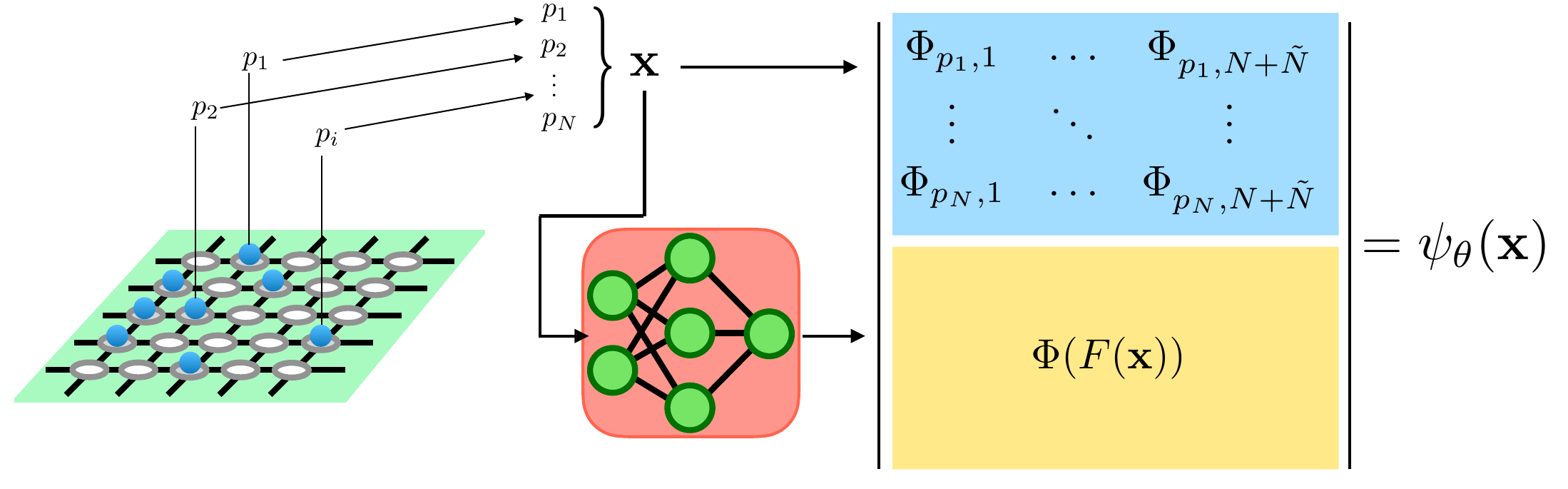}
    \caption{
        Diagram showing the evaluation of the hidden fermion determinant state with a fully marametrized constraint function. The left depicts an the electronic configuration in a lattice model $\xx$. The variational wave function amplitudes are obtained from the evaluation of an $(N + \tilde{N}) \times (N + \tilde{N})$ determinant. The first $N$ rows are obtained from slicing the matrix of orbitals $\Phi$ according to the electronic configurations $\xx$ (similar to a regular Slater determinant in Eq.~\ref{eq:Slater determinant}). In contrast to a regular determinant, $\Phi \in \mathcal{C}^{M \times (N + \tilde{N})}$. The bottom $\tilde{N}$ rows are obtained from the output of a permutation-invariant neural network of $\xx$. The output of the neural network must have $\tilde{N} \times(N \times \tilde{N})$ components. This representation of the bottom $\tilde{N}$ rows directly parametrized the composition of orbitals functions evaluated in the configurations of the hidden fermions given by the constraint function: $\Phi(F(\xx))$, which corresponds to the case when $\tilde{M} \rightarrow \infty$.
    }
    \label{fig:hidden fermions}
\end{figure*}

The hidden fermion formalism describes the $M$ physical degrees of freedom as the projection of an augmented Hilbert space with $M_\text{tot} = M + \tilde{M}$ degrees of freedom. The augmented Hilbert space is partitioned into $M$ \textit{visible} and $\tilde{M}$ \textit{hidden} degrees of freedom. The $M$ visible modes are a copy of the physical degrees of freedom. The projection is applied as follows: the occupancy of the visible modes $\xx$ matches the occupancy of the physical degrees of freedom, while the occupancy of the hidden modes $\tilde{\xx}$ by the $\tilde{N}$ hidden fermions is unequivocally determined by the occupancy of the visible modes. This correspondence can be summarized by a constraint function: $\tilde{\xx} = F(\xx)$, which must be permutation-invariant. The numbers of hidden modes and particles are hyper-parameters to be chosen by the user. The variational wave-function amplitudes are obtained as the overlap between a basis state and a parametrized Slater determinant in the augmented Hilbert space $\ket{\Psi_\text{SD}}$, characterized by $N + \tilde{N}$ orbital functions, projected into the physical subspace:
\begin{equation}
    \psi_\theta(\xx) = \bra{\xx, F(\xx)} \ket{\Psi_\text{SD}}.
\end{equation}
An extremely flexible wave function ansatz is obtained when the constraint function is parametrized $F \rightarrow F_\theta$ and when $\tilde{M} \rightarrow \infty$, even if $M$ is countable~\cite{robledo_moreno_fermionic_2022}. This ansatz can represent Slater-Jastrow states, as well as backflow wave functions. Furthermore, it can be explicitly spanned as the linear combination of all possible single, double, ..., $\tilde{N}$-tuple particle excitations from the Hartree-Fock ground state to the $\tilde{N}$ lowest unoccupied orbitals in the Hartree-Fock basis. Furthermore, the wave function amplitude evaluation in this formalism is straightforward, as shown in Fig.~\ref{fig:hidden fermions}. 

This family of variational states has demonstrated high degrees of accuracy in the description of the ground state properties of the Hubbard model in square and cylinder-like geometries~\cite{robledo_moreno_fermionic_2022}, as well as in the fully-connected lattice~\cite{gauvinndiaye2023mott}. Furthermore, this variational ansatz class has received increasing interest for the description of interacting nucleons~\cite{Lovato_Hidden_nucleons_2022, Fore_Hidden_nucleons_2022, gnech2023_hidden_nucleons}.

\subsubsection{Single-particle basis considerations}
The accuracy in the description of fermionic wave-functions by classical approximations is strongly affected by the single-particle basis choice that is used to span the Hilbert space, within a basis set~\cite{choo_fermionic_2020, rath_framework_2023, RobledoMoreno2023basisrotations}. The single-particle basis can be transformed according to:
\begin{equation}
    \ket*{\widetilde{\phi} _\alpha} = \sum_{p} \Xi_{p \alpha} \ket{\phi_p},
\end{equation}
where $\Xi \in \mathbb{C}^{N_\text{orb} \times N_\text{orb}}$ is the unitary matrix describing the change of basis. This class of changes of bases are know as single-particle basis transformations. The functional form of the Hamiltonian in Eq.~\ref{eq:electrons discrete} remains the same, but the coefficients are transformed according to the tensor transformations:
\begin{equation}
\begin{gathered}
    h_{pq} \rightarrow \sum_{pq} h_{pq} \Xi_{p\alpha} \Xi_{q\beta}; \\
    h_{pqrs} \rightarrow \sum_{pqrs} h_{pqrs} \Xi_{p\alpha} \Xi_{q\beta} \Xi_{r\delta} \Xi_{s\gamma}.
\end{gathered}
\end{equation}

For example, when considering molecular systems in the Hartree-Fock (molecular orbital) basis, the ground state wave function amplitudes are dominated by the electronic configuration corresponding to the occupation of the $N$ Hartree-Fock orbitals of lowest energy, while the amplitudes for other configurations rapidly decay. This sparse structure can be hard to capture by non-determinant based states~\cite{rath_framework_2023, RobledoMoreno2023basisrotations}. The Markov-Chain sampling of sparse (in the space of electronic configurations) wave functions can require a large number of samples to converge. Furthermore, wave functions dominated by zero-amplitude components pose challenges when estimating $g_\mu$ and $S_{\mu \nu}$ in Eq.~\ref{eq:averages}~\cite{Sinibaldi2023}.

More generally, the issue lies on the lack of invariance under the reparametrization of the Hamiltonian given by single-particle basis transformations. To address this issue, the variational state $\ket{\psi_\theta}$ can be dressed by a parametrized Gaussian unitary operator~\cite{RobledoMoreno2023basisrotations}
\begin{equation}
    U(\kappa) = \exp \left( \sum_{p q}\kappa _{p q} a^\dagger_p a_q \right) \; ,
\end{equation}
yielding the variational state $\ket*{\psi_{\{\kappa, \theta \}}} = U(\kappa) \ket{\psi_\theta}$. Both $\kappa$ and $\theta$ variational parameters are treated on equal footing. Since the action of $U(\kappa)$ on a basis state $\ket{\xx}$ yields an anti-symmetric product state on a rotated single-particle basis~\cite{Thouless1960}, then the state $\ket*{\psi_{\{\kappa, \theta \}}}$ is invariant under single-particle basis transformations. It has been empirically demonstrated that the expressive power and optimization landscape are improved when a NQS is dressed by $U(\kappa)$ in a collection of molecular systems and the Hubbard model~\cite{RobledoMoreno2023basisrotations}.

\subsection{Unitary operations}
\label{sec:unitary}

\subsubsection{Real-time evolution}

In recent years, an increasing body of research has been dedicated to real-time evolution of neural-network quantum states. As described in Sec.~\ref{sec:tvmc}, time-stepping is performed by projecting the state transformed by the unitary propagator onto the variational manifold: $\psi _{\theta + \delta \theta} (\xx) = \bra{\xx} e^{-i H \delta t} \ket{\psi _\theta}$, for a general trial state $\psi _\theta$. The precise way this projection is performed has been a focus point of the community.

Authors in Refs.~\cite{Carleo2012, Carleo2017_2} focus on bosonic models on the lattice. As a concrete example, consider the prototypical Bose-Hubbard model
\begin{equation}
\label{eq:bose-hubbard}
    H = -t \sum _{\langle p, q \rangle} \left( b^\dagger _p b_q + b^\dagger _q b_p \right) + \frac{U}{2} \sum _p n_p (n_p - 1) \; ,
\end{equation}
where $b_i$ and $b ^\dagger _i$ are bosonic annihilation and creation operators, respectfully, and $n_i = b_i ^\dagger b_i$ is the number operator used to label computational basis states $\left\{ \ket{n_1, \ldots, n_N} \; : n_i \in \mathbbm{N} _0 \right\}$. The Hamiltonian in Eq.~\ref{eq:bose-hubbard} should be thought of as a direct bosonic counterpart of Eq.~\ref{eq:fermi-hubbard}.

For such a system of lattice bosons and an initial state $\ket{\psi(t=0)} = \ket{\psi _0}$, a simple Jastrow-type~\cite{Jastrow1955} wavefunction can be constructed as:
\begin{equation}
\label{eq:bose-hubbard-jastrow}
    \ket{\psi (t)} = \exp \left( \sum _\mu \theta _\mu (t) \; \mathcal{A} _\mu (n) \right) \ket{\psi _0}
\end{equation}
where operators $\mathcal{A} _\mu$ are usually chosen on physical grounds as important excitations of the initial state. Authors in Ref.~\cite{Carleo2012} choose two-site correlations $\mathcal{A} _\mu \rightarrow \mathcal{A} _k (n) = \sum _i n_i n_{i + k}$, successfully reproducing glassy quench dynamics of long-lived metastable states. Authors in Ref.~\cite{Verdel2021, Schmitt2018} use a variational state of the form outlined in Eq.~\ref{eq:bose-hubbard-jastrow} but guide their choices of $\mathcal{A} _\mu$ by the cumulant expansion of the underlying classical Hamiltonian. The same type of state has been used to suggest local gauge invariance as a mechanism for disorder-free localization~\cite{Karpov2021} in two dimensional spin models.

As described in Sec.~\ref{sec:optimization}, in practice the logarithm $\ln \psi _{\theta (t)} (\xx) = \ln \braket{\xx}{\psi (t) }$ is directly parametrized instead of the wavefunction itself. In mathematical language, it is then trivial to show that the trial state in Eq.~\ref{eq:bose-hubbard-jastrow} corresponds to a linear model with physically-motivated pairwise kernel functions~\cite{Hofmann2008, Dawid2022} $\mathcal{A} _\mu \rightarrow \mathcal{A} (n_i, n_j) $.

As an example of a more complex trial state for Fermi-Hubbard dynamics~\ref{eq:fermi-hubbard}, authors in Ref.~\cite{Ido2015} consider a symmetry-projected Gutzwiller-Slater-Jastrow-Backflow wavefunction. This scheme introduces parametrized \textit{kernels} into a Jastrow-type state outlined in Eq.~\ref{eq:bose-hubbard-jastrow}, reproducing real-time dependence of a range of physically relevant observables.

In contrast with hand-crafted and fine-tuned trial states designed to capture known dynamical processes, the advent of automatic differentiation and large-scale adoption of NQS opened the door to expressive \textit{black-box} wavefunctions. Apart from improvements to ground-state results discussed, these trial states can capture accurate time-evolution over a range of physical Hamiltonians.

Authors in Ref.~\cite{Schmitt2020, Schmitt2022} employ a deep feed-forward convolutional neural network (CNN) to explore the transverse-field Ising model
\begin{equation}
\label{eq:transverse-field-ising}
    H  = - J \sum _{\left\langle i, j \right\rangle} \sigma ^z _i \sigma ^z _j - h \sum _i \sigma ^x _i \; ,
\end{equation}
whose dynamics has recently been shown to be accessible in Rydberg atom experiments~\cite{Barredo2018, Labuhn2016, GuardadoSanchez2018} and superconducting quantum devices~\cite{Kim2023}. We note that the Hamiltonian in Eq.~\ref{eq:transverse-field-ising} is a special case of Eq.~\ref{eq:spin-hamiltonians}.

In Ref.~\cite{Schmitt2020}, large square lattice dynamics, up to $12 \times 12$, have been treated using direct integration of the t-VMC Eq.~\ref{eq:averages}. When using deeper models like CNNs, in contrast with Eq.~\ref{eq:bose-hubbard-jastrow}, regularizing the inverse of the $S$~matrix (Eq.~\ref{eq:averages}) becomes crucial. Ref.~\cite{Schmitt2020} employs a custom spectral inverse, allowing the underlying calculations to proceed to $\sim 10 \, J^{-1}$ using very thorough Monte Carlo sampling at each step. Collapse and revival oscillations of ferromagnetic order are observed. Universal near-critical dynamics have been explored in Ref.~\cite{Schmitt2022} within the context of quantum Kibble-Zurek mechanism~\cite{Zurek1985, Zurek1996} in two dimensions. More recently, experimentally observable spectral functions have also been extracted from related t-VMC calculations~\cite{Mendes2023}. Disordered long-time dynamics of the Hamiltonian in Eq.~\ref{eq:transverse-field-ising} have been studied in Ref.~\cite{Burau2021}, including local unitary transformations in the NQS in a way similar to Sec.~\ref{sec:overlap_optimization}.

Generally, noise in estimators of matrix elements in Eq.~\ref{eq:averages} can conspire to render the $S$~matrix singular by making small eigenvalues vanish. Therefore, quickly and efficiently obtaining many uncorrelated samples from $\left| \psi \right| ^2$ is crucial. In addition, neural-network wavefunctions are usually overparametrized by design. While considered beneficial in AI applications~\cite{LeCun2015}, this feature can produce linearly dependent or vanishing rows and columns in $S$. Therefore, choosing an efficiently parameterized state is equally important. This leads to a superficially counter-intuitive observation -- adding more parameters to a trial state can reduce accuracy by making $S$ ill-conditioned.

Authors in Ref.~\cite{Medvidovic2023} perform a large simulation of the quantum rotor model
\begin{equation}
\label{eq:rotors}
    H = - \frac{g J}{2} \sum _k \pdv[2]{\phi _k} - J \sum _{\langle k, l \rangle} \cos(\phi _k - \phi _l) \; ,
\end{equation}
describing a system of continuous planar rotors on a lattice. An advantage of the NQS treatment is that one can work directly in a continuous basis $\ket{\bm{\phi}} = \ket{\phi _1, \ldots, \phi _N}$ of individual rotor angles $\phi _i$, bypassing discretization or local basis truncation that bias similar tensor-network approaches. The expression in Eq.~\ref{eq:rotors} is an effective Hamiltonian that captures the physics of superconducting Josephson junctions~\cite{Josephson1962, Josephson1965, Martinoli2000, Vogt2015, Kockum2019} that can be used to model phenomena related to superconducting or transmon qubit systems~\cite{trebst22}.

Using a Hamiltonian Monte Carlo sampler introduced in Sec.~\ref{sec:sampling}, calculations in Ref.~\cite{Medvidovic2023} reduce the sampling overhead encountered in previous works while maintaining good accuracy. In addition, the $S$~matrix inverse is regularized by a simple adaptive spectral regularization introduced in Eq.~\ref{eq:regularization} with the weight function
\begin{equation}
    w(\sigma ^2) = \left[ 1 + \left( \nicefrac{\lambda ^2}{\sigma ^2} \right) ^6 \right] ^{-1}
\end{equation}
where $\lambda ^2$ is set automatically at each step as a function of the spectrum $\left\{ \sigma ^2 _1, \ldots, \sigma ^2 _P \right\}$. This setup mitigates the underlying numerical instability for the CNN trial wavefunction, allowing accurate simulations up to $\sim 10 \, J^{-1}$ for $8 \times 8$ lattices and access to nontrivial observables such as the Loschmidt echo and the vorticity.

Due to the inversion problem described above, several alternatives to direct integration of the top-level differential equation $\dot{\theta} = -S^{-1} f$ have been proposed. Authors in Refs.~\cite{Gutierrez2022, Sinibaldi2023, Donatella2023} choose to completely replace the troublesome pseudoinverse with optimization subproblems at each step in real time. Specifically, Ref.~\cite{Sinibaldi2023} formulates this \textit{projection} step as a minimization of the quantum infidelity (see Sec.~\ref{sec:overlap_optimization})
\begin{equation}
\label{eq:ptvmc}
    \theta (t + \delta t) = \argmin _{\theta '} \mathcal{D} \left( \ket{\psi _{\theta '}} , e^{-i H \delta t} \ket*{\psi _{\theta (t)}}  \right)
\end{equation}
between the exact propagation of the state at time $t$ and its projection onto the variational manifold. Apart from avoiding numerical problems, authors report that this framework removes the bias appearing in estimators in Eq.~\ref{eq:averages} and Eq.~\ref{eq:imag_grad} when $\psi (\xx) \approx 0$ (see Sec.~\ref{sec:optimization}). The key observation is that the gradients of the infidelity objective in Eq.~\ref{eq:ptvmc} can be efficiently optimized using canned optimizers like Adam~\cite{Kingma2014} if one exploits the Trotter approximation
\begin{equation}
    e^{-i H \delta t} = e^{-i (H_0 + H_I) \delta t} \approx e^{-i H_0 \frac{\delta t}{2}} e^{-i H_I \delta t} e^{-i H_0 \frac{\delta t}{2}}
\end{equation}
in Eq.~\ref{eq:ptvmc}, assuming that the system Hamiltonian $H = H_0 + H_I$ decomposes into a term local in the computational basis $H_0$ and an interaction term $H_I$. Eliminating numerical instabilities, determining the scaling and reducing the computational cost of similar projective methods is an area of open research.

\subsubsection{Quantum circuit simulation}

\begin{figure}
    \begin{adjustbox}{width=\linewidth, center}
    \centering
    \begin{quantikz}
        \lstick[wires=5, label style={xshift=-12pt, anchor=center, rotate=90}]{\Large $\ket{\psi _{\theta =0}} \propto \ket{+}$} & \gate[wires=5, style={fill=blue!15}]{e^{-i \gamma _1 \mathcal{C}}} \gategroup[wires=5,steps=1,style={draw=none, dashed, rounded corners, inner xsep=2pt}, background,label style={label position=below, anchor=north, yshift=-0.2cm}]{Exact} & \gate[style={fill=yellow!20}]{e^{-i \beta_1 X}} \gategroup[wires=5,steps=1,style={dashed, rounded corners, inner xsep=2pt}, background,label style={label position=below, anchor=north, yshift=-0.2cm}]{Approx.} & \gate[wires=5, style={fill=blue!15}]{e^{-i \gamma _2 \mathcal{C}}} \gategroup[wires=5,steps=1,style={draw=none, dashed, rounded corners, inner xsep=2pt}, background,label style={label position=below, anchor=north, yshift=-0.2cm}]{Exact} & \gate[style={fill=yellow!20}]{e^{-i \beta_2 X}}
        \gategroup[wires=5,steps=1,style={dashed, rounded corners, inner xsep=2pt}, background, label style={label position=below, anchor=north, yshift=-0.2cm}]{Approx.} & \ \ldots\ \qw & \gate[wires=5, style={fill=blue!15}]{e^{-i \gamma _p \mathcal{C}}} \gategroup[wires=5,steps=1,style={draw=none, dashed, rounded corners, inner xsep=2pt}, background,label style={label position=below, anchor=north, yshift=-0.2cm}]{Exact} & \gate[style={fill=yellow!20}]{e^{-i \beta_p X}}
        \gategroup[wires=5,steps=1,style={dashed, rounded corners, inner xsep=2pt}, background, label style={label position=below, anchor=north, yshift=-0.2cm}]{Approx.} & \rstick[wires=5, label style={xshift=12pt, anchor=center, rotate=90}]{\Large $\ket{\psi _{\theta _\text{opt}}} \propto \ket{\bm{\gamma}, \bm{\beta}}$} \qw \\
        & & \gate[style={fill=yellow!20}]{e^{-i \beta_1 X}} & & \gate[style={fill=yellow!20}]{e^{-i \beta_2 X}} & \ \ldots\ \qw & & \gate[style={fill=yellow!20}]{e^{-i \beta_p X}} & \qw & \\
        & & \gate[style={fill=yellow!20}]{e^{-i \beta_1 X}} & & \gate[style={fill=yellow!20}]{e^{-i \beta_2 X}} & \ \ldots\ \qw & & \gate[style={fill=yellow!20}]{e^{-i \beta_p X}} & \qw & \\
        & & \gate[style={fill=yellow!20}]{e^{-i \beta_1 X}} & & \gate[style={fill=yellow!20}]{e^{-i \beta_2 X}} & \ \ldots\ \qw & & \gate[style={fill=yellow!20}]{e^{-i \beta_p X}} & \qw & \\
        & & \gate[style={fill=yellow!20}]{e^{-i \beta_1 X}} & & \gate[style={fill=yellow!20}]{e^{-i \beta_2 X}} & \ \ldots\ \qw & & \gate[style={fill=yellow!20}]{e^{-i \beta_p X}} & \qw & 
    \end{quantikz}
    \end{adjustbox}
    \caption{
        An outline of an algorithm using an RBM~(Eq.~\ref{eq:rbm}) trial state to simulate the quantum approximate optimization algorithm in Ref.~\cite{Medvidovic2020}. Interleaving layers of exact and approximate gates are applied to the RBM state, incurring a small but nonzero fidelity optimization error in each layer.
    }
    \label{fig:circuit}
\end{figure}
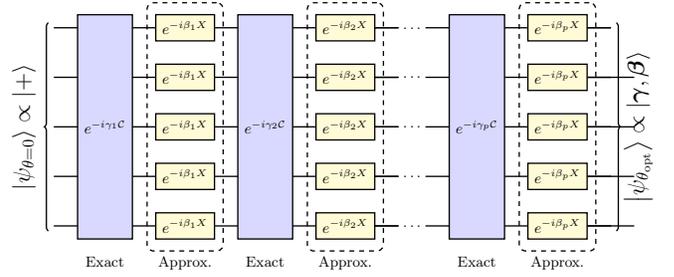

Quantum circuits offer a different perspective on unitary operations to NQS. In principle, arbitrarily complex unitary operators can be applied to $\ket{\psi _\theta}$ if one has a way of implementing a set of universal quantum logical gates~\cite{Nielsen2012}.

Authors in Ref.~\cite{Jonsson2018} choose the following one-qubit gates: $\textrm{RZ}(\phi) = \diag (1, e^{i \phi})$,
\begin{equation}
    \textrm{H} = \frac{1}{\sqrt{2}} \matrixquantity(1 & 1 \\ 1 & -1)
\end{equation}
and the two-qubit gate $\textrm{CZ} (\phi) = \diag(1, 1, 1, e^{i \phi})$. In that case, an analytical property of RBMs can be exploited -- both diagonal gates can be applied exactly (with fidelity $1$) by direct substitution rules in parameter space.

However, implementing the remaining Hadamard gate $\textrm{H}$ is nontrivial. The key observation is that any unitary $\mathcal{U}$ can be approximated by minimizing the infidelity $\mathcal{D}(\psi _{\theta '}, \mathcal{U} \psi _\theta)$ with respect to $\theta '$. Setting $\mathcal{U} \rightarrow \textrm{H}$, one obtains a recipe for approximating the state $\textrm{H} \ket{\psi _\theta}$ for any $\theta$. Calculations in Refs.~\cite{Jonsson2018, Medvidovic2020} show that this scheme can be used to obtain excellent approximations of Hadamard and Fourier transforms of entangled initial states.

Authors in Ref.~\cite{Medvidovic2020} show that upgrading the optimizer to SR within the framework of Eq.~\ref{eq:effective_hamiltonian} allows one to simulate deep quantum circuits. They focus on the quantum approximate optimization algorithm (QAOA)~\cite{Farhi2014, Farhi2016}, whose structure is outlined on Fig.~\ref{fig:circuit}. Taking advantage of exact gate application, the simple computation in Ref.~\cite{Medvidovic2020} successfully reaches 54 qubits at $p=4$ QAOA layers, approximately implementing 324 ZZ-rotation gates and 216 X-rotation gates without large-scale computational resources.

\subsection{Quantum state tomography and reconstruction}
\label{sec:tomography}

In contrast to the circuit simulation case in which the target state is known at each step, reconstructing a neural-network quantum state with only measurement data has been an open problem since the seminal work in Ref.~\cite{Torlai2018Tomography}. Exploiting the mathematical machinery of Sec.~\ref{sec:mle}, authors in Refs.~\cite{Torlai2018Tomography, Torlai2019TomographyRydberg, TorlaiMelko2018TographyDensityMatrix} directly minimize the many-basis KL-divergence cost in Eq.~\ref{eq:tomography_cost}. Restricted Boltzmann machines were (Eq.\ref{eq:rbm}) were used to capture both pure state and density matrix representations of reconstructed states.

Data-driven enhancements of VMC approach of the Rydberg Hamiltonian have been considered as well in Ref.~\cite{Czischek2022, moss2023enhancing}, showing that any amount of experimental measurement data can be used to efficiently \textit{pretrain} an RNN model, drastically reducing computation time.

Finally, authors in Ref.~\cite{Bennewitz2022} show that NQS based on the transformer~\cite{Vaswani2017} architecture can be trained on data generated by a variational quantum algorithm to enhance accuracy and decrease error bars.

\section{Concluding remarks and outlook}
\label{sec:conclusion}

Neural network parametrization of quantum states is a promising avenue to study systems of interacting quantum particles. The outstanding expressive power of this family if variational wave functions has allowed to obtain results which are competitive with other \textit{state-of-the-art} quantum many-body methods in challenging many-body systems.

There are a number challenges that this variational family faces, most of which are related to the optimization of the large number of variational parameters they posses. The first challenge is the calculation and inversion of the $S$ matrix for large models, which may become expensive to realize and numerically unstable to invert. A solution has been proposed to greatly reduce the cost of evaluating and inverting $S$~\cite{Chen2023}. NQS rely on gradients to optimize the wave function parameters. Another difficulty arises when the support of the wave function does not coincide with the support of the gradients. In this case, the Monte Carlo estimator for the gradients and the components of $S$ contains a bias~\cite{Sinibaldi2023}. While the bias for gradients can be removed, the bias for $S$ remains an open problem~\cite{Sinibaldi2023}. A solution based on overlap optimization is proposed in~\cite{Sinibaldi2023, kochkov2018variational}. Generally, there is lack in the understanding of the characteristics of the optimization landscape of NQS parametrizations. Initial progress has been made in its characterization using  neural tangent kernel techniques in the limit of infinite width NQS~\cite{Luo_2023NTK}. However, there are still open questions in practical implementations with finite numbers of parameters. 

Another challenge that these architectures face is the study of realistic systems of interacting electrons in second quantization. The number of terms in the Hamiltonian in Eq.~\ref{eq:electrons discrete} grows with the fourth power of the number of orbitals. The cost of evaluating the local energy for a moderate number of samples becomes prohibitively expensive, thus limiting the sizes of the basis sets that can be used for the simulation of realistic fermionic systems. Thus far, this has prevented any kind of extrapolation to the complete basis-set limit, which is possible in other many-body methods. A possible solution is to use the inductive bias where a the wave function amplitudes for a large number of electronic configurations is set to zero (active space calculation). In this case the choice of the single particle basis strongly affects the success of the highly compact wave function representation. Ref.~\cite{RobledoMoreno2023basisrotations} demonstrates the first active space NQS calculation with orbital optimizations, thus paving the path towards the study of more realistic fermionic systems using active spaces beyond the scope of exact diagonalization.  While MPS in conjunction with DMRG can be used to represent the wave function in the active space~\cite{zgid_density_2008, ghosh_orbital_2008, wouters_density_2014, ma_second-order_2017, wu_disentangling_2022}, this wave function representation is limited by the amount of entanglement entropy that it can sustain. 

Efficient real-time calculations also remain a challenge. With significant advances in neural-network architectures~\cite{Dawid2022, Schmitt2020} and Monte-Carlo sampling~\cite{Medvidovic2023}, larger systems can be simulated for a longer time, reaching experimentally relevant system sizes and time scales in some cases. However, scaling the t-VMC approach to even larger systems and NQS will likely require rethinking the method itself -- the QGT matrix $S$ inverse is a clear bottleneck. Alternative approaches have been proposed~\cite{Sinibaldi2023, Luo2020, Dugan2023} and scaling them up is an area of active research.

Neural network quantum states remain a powerful way of compressing complex many-body quantum physics. As model architectures, optimization algorithms and researchers' intuition evolve, these methods are entering the domain of practical application, especially in quantum chemistry and benchmarking near-term quantum hardware.

\section*{Acknowledgements}
M.~M. and J.~R.~M acknowledge support from the CCQ graduate fellowship in computational quantum physics. The Flatiron Institute is a division of the Simons Foundation. The authors acknowledge useful discussions with Antoine Georges, Christopher Roth, Schuyler Moss, Agnes Valenti, Alev Orfi, Anna Dawid and Anirvan Sengupta.

\section*{Data availability statement}
There is no data associated with the manuscript.

\bibliographystyle{naturemag}
\bibliography{references}

\onecolumngrid
\appendix

\section{Derivation of the projected equations of motion}
\label{appendix:vmc-eom}

In Eq.~\ref{eq:param_update} in the main text, a simple update scheme for variational parameters is presented:
\begin{equation}
\label{eq:param-update-normalized}
    \ket{\Psi _{\theta + \delta \theta}} = e^{-\delta \tau H} \ket{\Psi _\theta}
\end{equation}
where $\ket{\Psi _\theta} = \ket{\psi} / \sqrt{\braket{\psi}}$ is the normalized version of our variational state of choice, turning the proportionality sign in Eq.~\ref{eq:param_update} into an equality. We note that the differential version of Eq.~\ref{eq:param-update-normalized}, $\frac{\dd}{\dd \tau} \ket{\Psi _\theta} = - H \ket{\Psi _\theta}$ is known as the Bloch equation and can serve as an equivalent starting point for the following analysis.

In this appendix, we derive the projected equations of motion given in Eq.~\ref{eq:averages} for variational parameters $\theta$. We note that the following derivation plays out almost identically for the case of real-time evolution, upon substitution $\delta \tau \rightarrow i \delta t$. Borrowing notational conventions set up in Sec.~\ref{sec:vmc}, we begin by Taylor-expanding both sides of Eq.~\ref{eq:param_update} in small parameters:
\begin{equation}
\label{eq:vmc-taylor-exp}
\begin{gathered}
    e ^{-\delta \tau H} \ket{\Psi _{\theta}} = \left( \mathbbm{1} - \delta \tau H \right) \ket{\Psi _{\theta}} + \cdots \\
    \ket{\Psi _{\theta + \delta \theta}} = \ket{\Psi _{\theta}} + \sum _\mu \delta \theta ^\mu \partial _\mu \ket{\Psi _{\theta}} = \left( \mathbbm{1} + \sum _\mu \delta \theta ^\mu \mathcal{D} _\mu \right) \ket{\Psi _{\theta}} + \cdots \, .
\end{gathered}
\end{equation}
where we exploit the following chain of identities
\begin{equation}
    \partial _\mu \ket{\Psi _{\theta}} = \sum _\xx \partial _\mu \Psi _\theta (\xx) \ket{\xx} = \sum _\xx \Psi _\theta (\xx) \, \partial _\mu \ln \Psi _\theta (\xx) \ket{\xx} = \left( \sum _{\xx '} \partial _\mu \ln \Psi _\theta (\xx) \ketbra{\xx '} \right) \left( \sum _\xx \Psi _\theta (\xx ) \ket{\xx} \right) = \mathcal{D} _\mu \ket{\psi _\theta }
\end{equation}
to construct a convenient representation of derivative operators
\begin{equation}
    \partial _\mu \mapsto \mathcal{D} _\mu = \sum _{\xx} \partial _\mu \ln \Psi _\theta (\xx) \ketbra{\xx}
\end{equation}
on the variational manifold $\mathcal{M} _\psi$. Truncating the Taylor expansions in Eq.~\ref{eq:vmc-taylor-exp}, we get
\begin{equation}
    \sum _\mu \mathcal{D} _\mu \ket{\Psi _\theta} \dot{\theta} ^\mu = -H \ket{\Psi _\theta}
\end{equation}
and reduce the original equation to finding $\dot{\theta} ^\mu = \argmin \norm{\sum _\mu \mathcal{D} _\mu \dot{\theta} ^\mu - H}$. To build a convenient way of solving for $\dot{\theta}$, we left-multiply by $ \bra{\Psi _\theta} \mathcal{D} ^\dagger _\nu$ and relabel indices:
\begin{equation}
\label{eq:normalized-qgt}
    \sum _\nu \ev{ \mathcal{D} _\mu ^\dagger \mathcal{D} _\nu} _{\Psi _\theta} \dot{\theta} ^\nu = - \ev{ \mathcal{D} _\mu ^\dagger H} _{\Psi _\theta} \; ,
\end{equation}
yielding a linear system.

Finally, we need to unpack the fact that we usually do not have access to the full normalized state $\ket{\Psi}$. Instead, we need to substitute $\ket{\Psi} \mapsto \ket{\psi} / \sqrt{\braket{\psi}}$. The expectation value changes trivially into the form that can be evaluated using Monte Carlo sampling, as discussed in the main text. However, variational derivative operators $\mathcal{D} _\mu$ pick up an extra term because
\begin{equation}
    \partial _\mu \ln \Psi = 
    \partial _\mu \ln \psi _\theta - \frac{1}{2} \frac{\braket{\partial _\mu \psi _\theta}{\psi _\theta} + \braket{\psi _\theta}{\partial _\mu \psi _\theta}}{\braket{\psi _\theta}} = 
    \partial _\mu \ln \psi _\theta - \Re \ev{\OO _\mu} _{\psi _\theta}
\end{equation}
assuming that $\ket{\partial _\mu \psi _\theta} = \partial _\mu \ket{\psi _\theta}$ and that all of the variational parameters $\theta$ are real: $\theta \in \mathbbm{R} ^P$. Therefore, we are free to substitute $\mathcal{D}_\mu \mapsto \OO _\mu - \Re \ev{\OO _\mu}$ in Eq.~\ref{eq:normalized-qgt}, resulting in
\begin{equation}
    \Re \sum _\nu \left\{ \ev{ \OO _\mu ^\dagger \OO _\nu} _{\psi _\theta} - \ev{ \OO _\mu ^\dagger} _{\psi _\theta} \ev{\OO _\nu} _{\psi _\theta} \right\} \dot{\theta} ^\nu = - \Re \left\{ \ev{ \OO _\mu ^\dagger H} _{\Psi _\theta} - \ev{ \OO _\mu ^\dagger} _{\psi _\theta} \ev{H} _{\psi _\theta} \right\} \; ,
\end{equation}
which is identical to expressions in Eq.~\ref{eq:averages} in the main text.

\end{document}